\documentclass[structabstract]{aa}

\usepackage{graphicx}
\usepackage{txfonts}
\usepackage{natbib}
\bibpunct{(}{)}{;}{a}{}{,} 

\newcommand{\modified}[1]{#1}

\begin{document}

\title{Wide field polarimetry around the Perseus cluster at 350~MHz}
\titlerunning{Perseus polarimetry at 350~MHz}
\author{M.A. Brentjens}
\authorrunning{Brentjens}
\institute{ASTRON, P.O. Box 2, 7990 AA Dwingeloo, the Netherlands}
\offprints{brentjens@astron.nl}
\date{Received 1 July 2010 / Accepted 28 October 2010}

\abstract{}
{This paper investigates the fascinating diffuse polarization
  structures at 350 MHz that have previously been tentatively
  attributed to the Perseus cluster and, more specifically, tries to
  find out whether the structures are located at (or near) the Perseus
  cluster, or in the Milky Way.}
{A wide field, eight point Westerbork Synthesis Radio Telescope mosaic
  of the area around the Perseus cluster was observed in full
  polarization. The frequency range was 324 to 378~MHz and the
  resolution of the polarization maps was
  $2\arcmin\times3\arcmin$. The maps were processed using Faraday
  rotation measure synthesis to counter bandwidth depolarization. The
  RM-cube covers Faraday depths of $-384$ to $+381$~rad~m$^{-2}$ in
  steps of $3$~rad~m$^{-2}$.}
{ There is emission all over the field at Faraday depths between $-50$
  and $+100$~rad~m$^{-2}$. All previously observed structures were
  detected. However, no compelling evidence was found supporting
  association of those structures with either the Perseus cluster or
  large scale structure formation gas flows in the Perseus-Pisces
  super cluster. On the contrary, one of the structures is clearly
  associated with a Galactic depolarization canal at 1.41~GHz. Another
  large structure in polarized intensity, as well as Faraday depth at
  a Faraday depth of $+30$~rad~m$^{-2}$, coincides with a dark object
  in WHAM H$\alpha$ maps at a kinematic distance of
  $0.5\pm0.5$~kpc. All diffuse polarized emission at 350~MHz towards
  the Perseus cluster is most likely located within 1~kpc from the
  Sun. The layers that emit the polarized radiation are less than
  40~pc/$\|\vec{B}_\parallel\|$ thick.  }
{}


\maketitle

\section{Introduction}

De Bruyn \& Brentjens (2005) discovered \modified{fascinating
  structures in sensitive 350~MHz polarimetric observations of the
  \object{Perseus cluster} (\object{Abell~426}) with the Westerbork
  Synthesis Radio Telescope (WSRT). These structures were discovered
  using a novel data reduction procedure called \emph{RM-synthesis}
  \citep{BrentjensDeBruyn2005}, which extends the work of
  \citet{Burn1966} to multiple lines of sight and arbitrary source
  spectra, and were tentatively attributed to the Perseus cluster}.
More information on Faraday depth and depolarization can be found
elsewhere in the literature
\citep{Tribble1991,SokoloffEtAl1998,Vallee1980}.



There were two classes of polarized emission: \modified{clearly Galactic, more
or less uniform emission at $\phi=0$ to $+12$~rad~m$^{-2}$}, and
distinct features that appeared at Faraday depths between +20 and
+80~rad~m$^{-2}$. The second class consisted of two straight features
and three other distinct structures. The straight features were the
``front'' on the western side of the field and the ``bar'' at the
northern edge. A lenticular feature, partially embedded in the front
with its major axis parallel to the front, was called the
``lens''. Directly east of the lens was a very bright, shell-like
object called the ``doughnut''. A patch of polarized emission north of
the extended tail of \object{NGC~1265} was called the ``blob''.

Two possible locations were considered for the second class of
emission: our Galaxy, in particular the Perseus arm, and the Perseus
cluster itself. We favoured the latter \modified{because}
\begin{itemize}
\item there was a gap in Faraday depth between the large scale
Galactic foreground and the distinct features;
\item the typical scales in both polarized intensity and $Q$ and $U$
of the features were considerably smaller than the low $\phi$
emission;
\item higher Faraday depths appeared to occur closer to
\object{3C~84};
\item the largest structure, the front, was located in the direction
of the interface between the Perseus-Pisces super cluster filament and
the Perseus cluster;
\item a mini survey of 11 polarized point sources within a few degrees
of \object{3C~84} suggested an excess in Faraday depth of
$+40$ to $+50$~rad~m$^{-2}$ of the emission with respect to these
background sources, which was difficult to explain by a small Galactic
Faraday rotating cloud.
\end{itemize}

If the objects indeed resided at or near the cluster, the
``front'' could be a large scale structure formation shock at the
outskirts of the Perseus cluster, squashing a buoyant bubble (the
``lens''). It was suggested that the ``doughnut'' and ``blob'' are
bubbles that were released more recently into the cluster medium by
AGN. The discovery of X-ray cavities
\citep{FabianEtAl2003ShocksAndRipples,ClarkeEtAl2004} much closer to
\object{3C~84}, combined with simulation work on buoyant bubbles and
radio relic sources
\citep{EnsslinEtAl1998,EnsslinGopalKrishna2001,Brueggen2003,
  EnsslinBrueggen2002} reinforced this idea. Highly polarized relic
sources have been observed in several galaxy clusters
\citep{RottgeringEtAl1997,EnsslinEtAl1998,GovoniEtAl2001,GovoniEtAl2005},
but never in the \object{Perseus cluster}.

A Galactic origin for the high $\phi$ structures could nevertheless not
be ruled out, the main issue being that none of the structures had a
counterpart in Stokes $I$. Because the noise in $I$ is considerably
higher than the noise in $Q$ and $U$ due to classical source
confusion, \modified{a counterpart was only expected for the brightest
  polarized structures}. It was nevertheless puzzling that not even
the ``doughnut'' was detected in total intensity, although its
polarized surface brightness is only slightly lower than the noise
level in Stokes $I$. One possible explanation is that the Stokes $I$
surface brightness is intrinsically low.  This requires a fractional
polarization close to the theoretical limit for a synchrotron emitting
plasma with an isotropic distribution of the electron velocity vectors
\citep[$\approx$70\%, see e.g.  ][]{LeRoux1961,RybickyLightman}.

Another explanation is that the Stokes $I$ surface brightness is only
\emph{apparently} low. This is a well known property of
interferometric observations of Galactic synchrotron emission
\citep{Wieringa1993}, which is extremely smooth in Stokes $I$ and is
therefore not picked up by the shortest WSRT baseline of
$\lesssim40\lambda$.  The Stokes $Q$ and $U$ structure, however,
\emph{is} detectable at much longer baselines due to small scale
changes in the observed polarization angle. The apparent fractional
polarization can therefore far exceed 100\%. This effect could be
\modified{important in the observations by
  \citet{DeBruynBrentjens2005} because several observing sessions
  lacked the shortest spacing}.

In this paper I present an eight point WSRT mosaic of the region
around the Perseus cluster. Faraday rotation measure synthesis was
used to map polarized intensity in the area where
$2^\mathrm{h}58^\mathrm{m} \le \alpha \le 3^\mathrm{h}35^\mathrm{m}$
and $39\degr20\arcmin \le \delta \le 44\degr$ (J2000) for $-384 \le
\phi \le +381$~rad~m$^{-2}$. The primary goal is to assess whether the
structures previously observed by \citet{DeBruynBrentjens2005} are
located near the Perseus cluster or in the Milky Way.

An angular distance of one degree corresponds to 1.5~Mpc at the
distance of the \object{Perseus cluster} and 35~pc at the distance of
the Perseus \modified{arm ($\approx 2$~kpc)}. The redshift of the
Perseus cluster is $z=0.0167$ \citep{StrubleRood1999}. I assume that
$H_0 = 72\pm2\ \mbox{km}\ \mbox{s}^{-1}\ \mbox{Mpc}^{-1}$
\citep{SpergelEtAl2003}.


\section{Observations}

\begin{table}
\caption{Pointing positions of the fields.}
\label{brentjens_perseusmosaic_tbl:fields}
\begin{center}
\begin{tabular}{lll}
\hline
\hline
Name & $\alpha$ & $\delta$ \\
     & (J2000)          & (J2000) \\
\hline
A & $ 3^\mathrm{h}06^\mathrm{m}24^\mathrm{s}$ & $42\degr23\arcmin$ \\
B & $ 3^\mathrm{h}06^\mathrm{m}24^\mathrm{s}$ & $41\degr07\arcmin$ \\
C & $ 3^\mathrm{h}13^\mathrm{m}12^\mathrm{s}$ & $41\degr07\arcmin$ \\
D & $ 3^\mathrm{h}13^\mathrm{m}12^\mathrm{s}$ & $42\degr23\arcmin$ \\
E & $ 3^\mathrm{h}20^\mathrm{m}00^\mathrm{s}$ & $42\degr23\arcmin$ \\
F & $ 3^\mathrm{h}20^\mathrm{m}00^\mathrm{s}$ & $41\degr07\arcmin$ \\
G & $ 3^\mathrm{h}26^\mathrm{m}48^\mathrm{s}$ & $41\degr07\arcmin$ \\
H & $ 3^\mathrm{h}26^\mathrm{m}48^\mathrm{s}$ & $42\degr23\arcmin$ \\
\hline
\hline
\end{tabular}
\end{center}
\end{table}

The observations were conducted with the Westerbork Synthesis Radio
Telescope \citep{BaarsHooghoudt1974, DeBruynWSRT1996}.
The array consists of fourteen parallactic 25~m dishes on an east-west
baseline and uses earth rotation to fully synthesize the uv-plane in
12~h. \modified{There are ten fixed dishes (RT0~--~RT9) and four movable
telescopes (RTA~--~RTD)}.

\modified{The distance between two adjacent fixed telescopes is 144~m}.
The distances between the movable dishes were kept constant
(RTA~--~RTB = RTC~--~RTD = 72~m, RTB~--~RTC = 1224~m), while the
distance RT9~--~RTA was changed for every observing session. The
uv-plane is therefore sampled at regular intervals of 12~m out to the
longest baseline of 2760~m, lacking only the 0, 12, and 24~m spacings.
The regular interval causes an elliptical grating ring with an
east-west radius of $4\degr$ and a north-south radius of
$4\degr/\sin\delta$ at 350 MHz. At this frequency the $-5$~dB and
$-10$~dB points of the primary beam are at radii of $70\arcmin$\ and
$120\arcmin$\ respectively. The observations are sensitive to angular
scales up to $90\arcmin$\ at a resolution of
$74\arcsec\times96\arcsec$\ full width at half maximum (FWHM).

\modified{The observations were conducted in mosaic mode. The pointing centres
are listed in Table~\ref{brentjens_perseusmosaic_tbl:fields}.  Each
session began at a different field to improve the position angle
distribution in the uv-plane (see also
Table~\ref{brentjens_perseusmosaic_tbl:observations}). The dwell time
per pointing was 150~s and the total integration time after six
observing sessions was $8^\mathrm{h}22^\mathrm{m}$ per field.}

The eight frequency bands \modified{are} each 10~MHz wide and
\modified{are} centred at 319, 328, 337, 346, 355, 365, 374, and
383~MHz. The multi-frequency front ends \citep{Tan1991} of the WSRT
\modified{have} linearly polarized feeds for this frequency range. The $x$ dipole
is oriented east-west, \modified{the $y$ dipole north-south}. The correlator
produced 64 channels in all four cross correlations for each band with
an integration time of 10~s. The observations were using 180\degr\
front end phase switching. The on-line system applied a Hanning
\citep{Harris1978} lag-to-frequency taper, effectively halving the
frequency resolution.

The observations were bracketed by two pairs of calibrators, each
consisting of one polarized and one unpolarized source.
\object{3C~345} and \object{3C~48} were observed before the mosaic and
\object{3C~147} and the eastern hot spot of \object{DA~240}
afterwards.

\begin{table}
\caption{Basic information on the six 12 hour sessions.}
\label{brentjens_perseusmosaic_tbl:observations}
\begin{center}
\begin{tabular}{llll}
\hline
\hline
Obs. ID     & Obs. start          & RT9~--~RTA  & First field\\
            & (UTC)               & (m)         &            \\
\hline                           
10308301    & 2003/11/21 16:59:00 & 48        & H\\
10308473    & 2003/11/28 16:24:40 & 60        & E\\
10308694    & 2003/12/07 15:49:10 & 36        & F\\
10308727    & 2003/12/09 15:27:40 & 72        & B\\
10308798/99 & 2003/12/13 15:18:50 & 84        & D\\
10308817    & 2003/12/15 15:04:10 & 96        & A\\
\hline
\hline
\end{tabular}
\end{center}
\end{table}


\section{Data reduction}
\label{brentjens_perseusmosaic_sec:data_reduction}

Flagging, imaging, and self calibration were performed with the AIPS++
package \citep{McMullinGolapMyers2004}. Flux scale calibration,
polarization calibration, ionospheric Faraday rotation corrections,
and deconvolution were performed with a calibration package written by
the author and based on the table, measures, and fitting modules of
AIPS++/CASA. Channels and frequency bands are numbered \modified{from} 1.

\subsection{Data quality}
\label{brentjens_perseusmosaic_sec:data_quality}

\modified{Although the lowest and highest sub bands had to be discarded, the
data quality was generally good and interference levels were low}. The
Sun was still up at the beginning of all but the first observing sessions
(see Table~\ref{brentjens_perseusmosaic_tbl:observations}). \modified{The system
temperatures were usually between 130~K and 220~K with the median at
175~K. The expected thermal RMS image noise in a clean Hanning tapered
channel after $8^\mathrm{h}22^\mathrm{m}$ of integration is
2.6~mJy~beam$^{-1}$ \citep{ThompsonMoranSwenson}}.

\modified{Because of the Hanning tapering, I processed only the odd numbered
channels from 5 to 59 inclusive.  Approximately 20\% of the data in
these channels were flagged, hence the expected thermal RMS image
noise per field at full resolution is $0.22$~mJy~beam$^{-1}$ after
averaging all processed channel maps from the six usable spectral
windows.  The visibilities were time averaged to 30~s before
calibration and imaging to reduce processing time}.

\subsection{Calibration}
\label{brentjens_perseusmosaic_sec:calibration}

\modified{
The flux scale, bandpass, and polarization leakages were calibrated
simultaneously per individual channel by solving the
Hamaker-Bregman-Sault Measurement Equation \citep{HBS2} for the
unpolarized calibrator sources \object{3C~147} and \object{3C~48}.
The \citet{PerleyTaylor1999} calibrator fluxes, which extend the
\citet{BaarsEtAl1977} flux scale to lower frequencies}\footnote{The
  flux scale of WSRT observations has since 1985 been based on a
  325~MHz flux of 26.93~Jy for \object{3C~286} (the
  \citet{BaarsEtAl1977} value). On that flux scale, the 325~MHz flux
  of \object{3C~295} is 64.5~Jy, which is almost 7\% more than the
  value assumed at the VLA and in this paper (A.~G.~de~Bruyn, private
  communication).}, \modified{established the absolute flux scale}.

\modified{The polarization leakages were solved per channel because of their
strong 17~MHz semi-periodic frequency dependence \citep[see
e.g.][]{DeBruynBrentjens2005}. The diagonal phases of the RT0 Jones
matrix were fixed at 0~rad. The remaining $x$~--~$y$ phase difference
was determined using the polarized sources}.

\modified{The ionospheric Faraday rotation was corrected with the method from
\citet{Brentjens2008}. All fields were subsequently individually self
calibrated \citep{PearsonReadhead1984} with a CLEAN component
\citep{Hogbom1974} based sky model in $I$ and $Q$.  The strongly
frequency dependent polarization leakages required a separate CLEAN
model per channel}.

Fields A and B were calibrated with three phase-only iterations
because the total flux in these fields was too low for amplitude self
calibration. The remaining fields were calibrated with two phase-only
iterations and one amplitude/phase iteration. Each 10~MHz band was
self calibrated with a single Jones matrix per \modified{antenna at 30~s
intervals.}

\subsection{Imaging}
\label{brentjens_perseusmosaic_sec:imaging}

All fields were imaged and deconvolved separately.  The point spread
functions (PSFs) and dirty channel images in all Stokes parameters
were created using AIPS++. The uv-plane was uniformly weighted.
Because of a fractional bandwidth of 15\%, \modified{the maps had to be
convolved to a common resolution of $74\arcsec\times96\arcsec$~FWHM,
elongated north-south, using a Gaussian uv-plane taper}.  All maps are
in north celestial pole (NCP) projection with the projection centre at
\object{3C~84} (J2000: $\alpha=3^\mathrm{h}19^\mathrm{m}48\fs1601$,
$\delta= +41\degr30\arcmin42\farcs106$).  The dirty maps have
2048$\times$2048 pixels of $30\arcsec\times30\arcsec$ each.

\modified{The central 1024$\times$1024 pixels of the dirty images were
deconvolved using a H\"ogbom CLEAN \citep{Hogbom1974}. The CLEAN mask
consisted of all Stokes $I$ pixels brighter than 6, 5, 10, 10, 15, 8,
10, and 10~mJy~beam$^{-1}$ for fields A~--~H, respectively. The
deconvolution was stopped whenever the maximum residual in the masked
area was below 0.5~mJy~beam$^{-1}$ or when 10\,000 iterations were
completed without reaching the threshold. The resulting model images
were convolved with a $74\arcsec\times96\arcsec$~FWHM elliptical
Gaussian and added back to the residual images. The deconvolution of
the $Q$ and $U$ images was terminated after 10\,000 iterations or if a
threshold of 0.5~mJy~beam$^{-1}$ was reached}.

\modified{The primary beam corrected images were combined into one mosaic image
per channel per Stokes parameter. The restored Stokes $Q$ and $U$
mosaic maps were subsequently convolved to a resolution of $2\farcm
0\times3\farcm 0$ FWHM to enhance the signal to noise ratio of
extended emission. The expected RMS thermal noise after averaging all
low resolution images for one field is increased to
0.4~mJy~beam$^{-1}$ near the pointing centres and 0.3~mJy~beam$^{-1}$
in the areas surrounded by four pointings because the convolution
suppresses data from long baselines}.

\modified{Although the theoretical $Q$ and $U$ noise is approached at the
intersection of fields A, B, C, and D, the RMS image noise in most
areas of the mosaic is a factor of two higher due to Stokes $U$
dynamic range problems associated with \object{3C~84}}.

\subsection{RM-synthesis}
\label{brentjens_perseusmosaic_sec:rm_synthesis}

\begin{figure}
\centering
\includegraphics[width=\columnwidth]{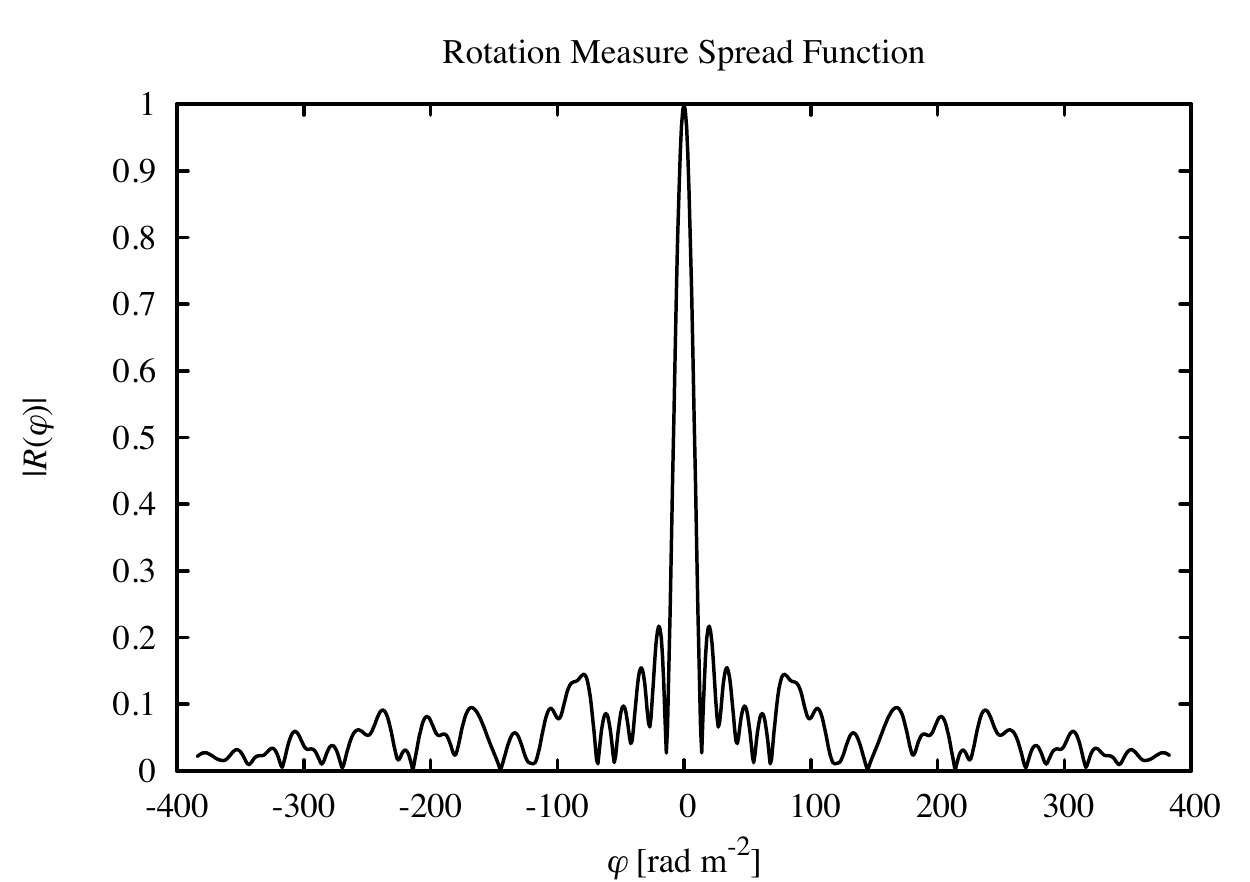}
\caption{Absolute value of the RMSF corresponding to the frequency
coverage of the observations. The FWHM of the main peak is
$16.4$~rad~m$^{-2}$.}
\label{brentjens_perseusmosaic_fig:rmsf}
\end{figure}

\begin{figure*}
\centering
\includegraphics[width=\textwidth]{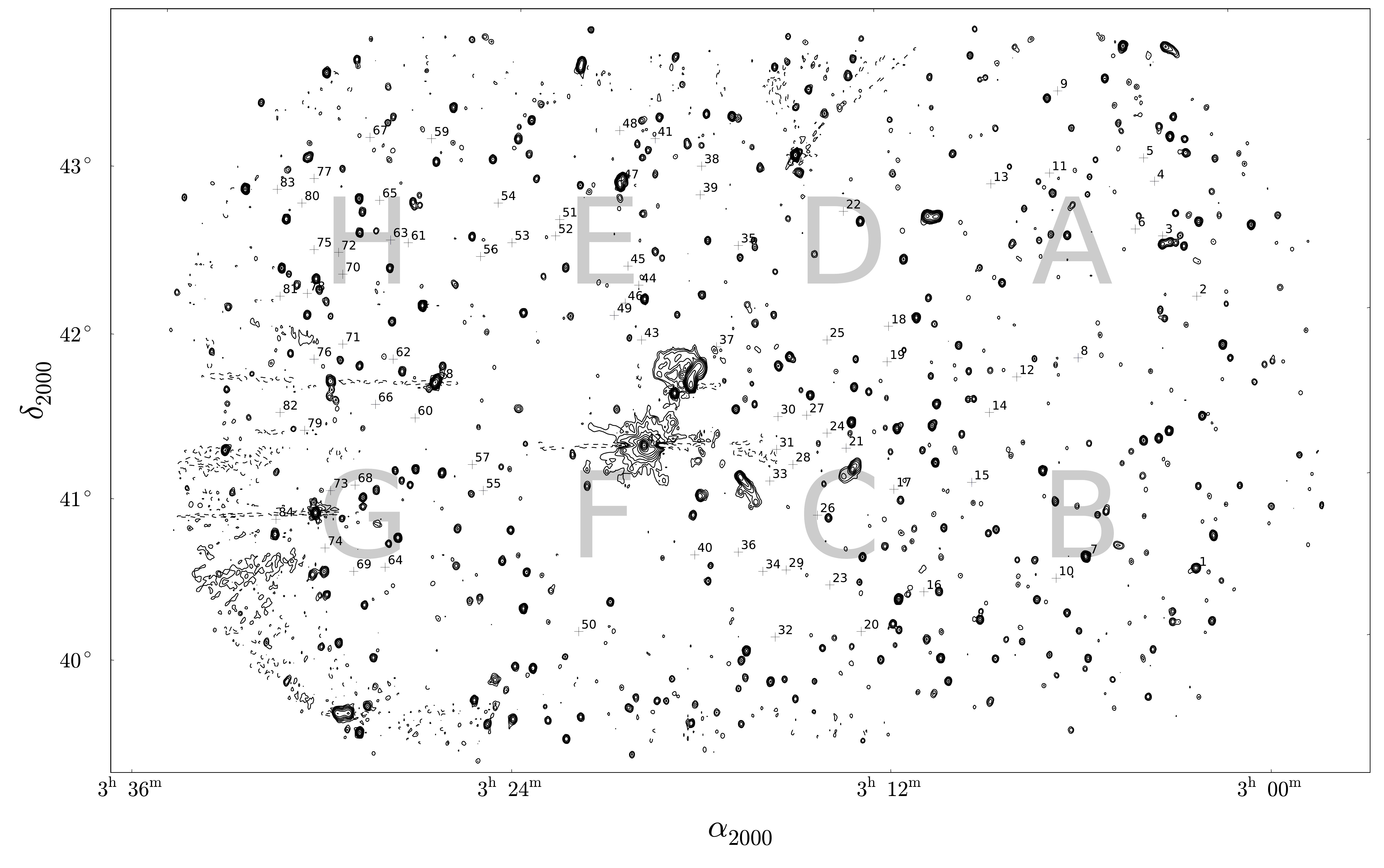}
\caption{Overview of the mosaic in total intensity at a resolution of
$74\arcsec\times96\arcsec$ FWHM. The contours are drawn at $-9$, $+9$,
18, 36~mJy~beam$^{-1}$~$\cdots$~18~Jy~beam$^{-1}$. The numbered
crosses mark the locations of the Faraday spectra in
Fig.~\ref{brentjens_perseusmosaic_fig:spectra-1}.
The dynamic range in this map is approximately 6000:1.}
\label{brentjens_perseusmosaic_fig:imap}
\end{figure*}

\modified{The 143 good quality polarization maps were processed using
RM-synthesis \citep{BrentjensDeBruyn2005} to avoid bandwidth
depolarization. The RM-cube covers the range $-384\leq \phi \leq
+381$~rad~m$^{-2}$ in steps of 3~rad~m$^{-2}$.  The absolute value of
the corresponding RMSF is displayed in
Fig.~\ref{brentjens_perseusmosaic_fig:rmsf}. The FWHM of the main peak
is 16.4~rad~m$^{-2}$ and the side lobes are of the order of
15\%~--~20\%, requiring deconvolution in $\phi$-space using an
RM-CLEAN}\footnote{\modified{This type of deconvolution works well if there is
  only one component along the line of sight, or if the emitting
  regions are separated in $\phi$ by more than the width of the
  RMSF. There may be systematic errors in the Faraday depth of each
  individual component up to the width of the RMSF if the components
  are closer together and of similar brightness (L.~Rudnick, private
  communication). This is effectively an interference effect due to
  the complex nature of the RMSF. The effect can potentially increase
  the scatter in $\phi$, but will average out over sufficiently many
  lines of sight. Although it is unimportant for the conclusions drawn
  in the remainder of this paper, it is something to take into account
  when doing RM work above 1~GHz, because the RMSF in such
  observations is typically very wide.}}  \modified{similar to the work of
\citet{HealdEtAl2009}.}


\section{Images}
\label{brentjens_perseusmosaic_sec:images}

\begin{figure*} \centering \begin{minipage}{0.48\textwidth}
   \includegraphics[width=\textwidth]{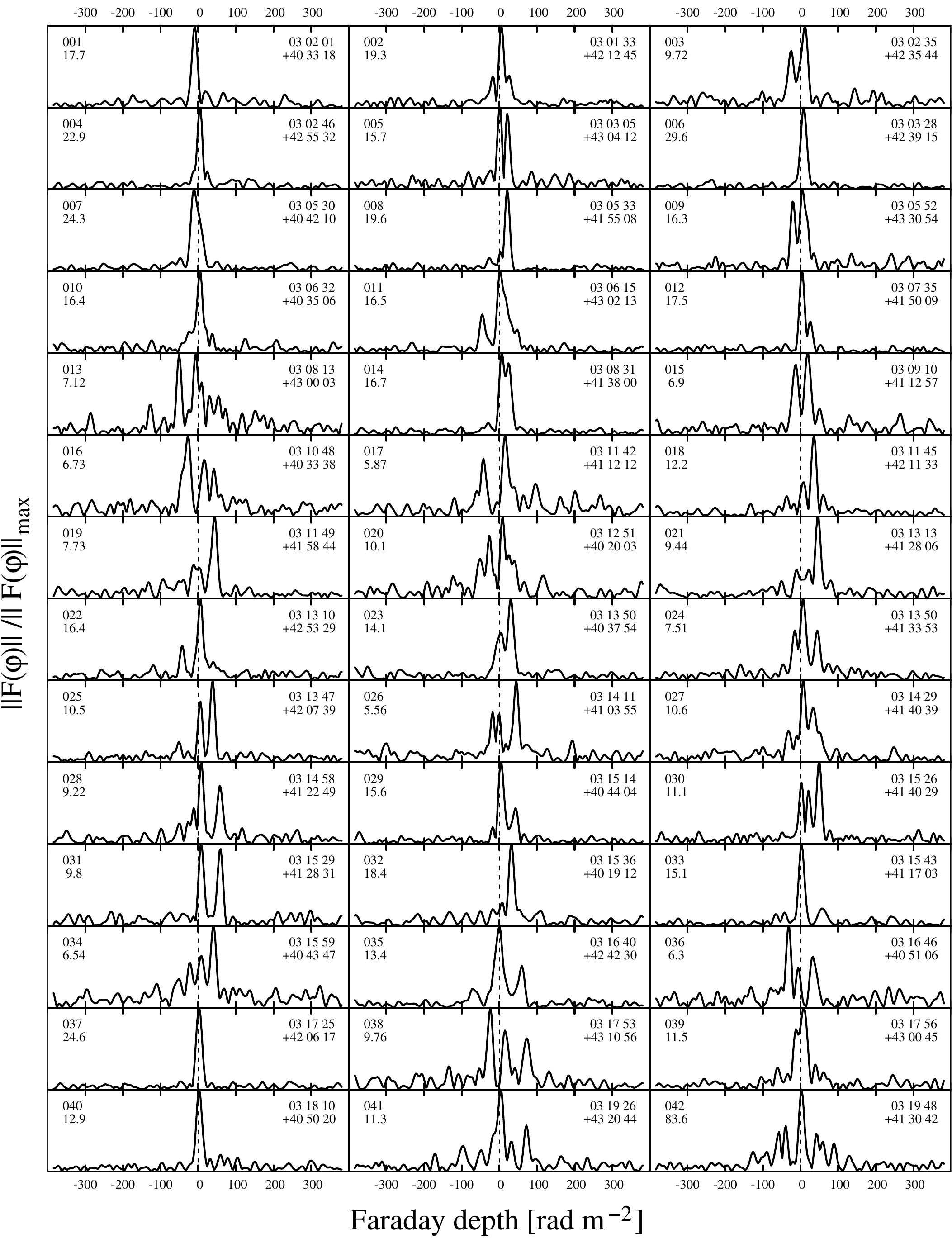}
 \end{minipage} \begin{minipage}{0.48\textwidth}
   \includegraphics[width=\textwidth]{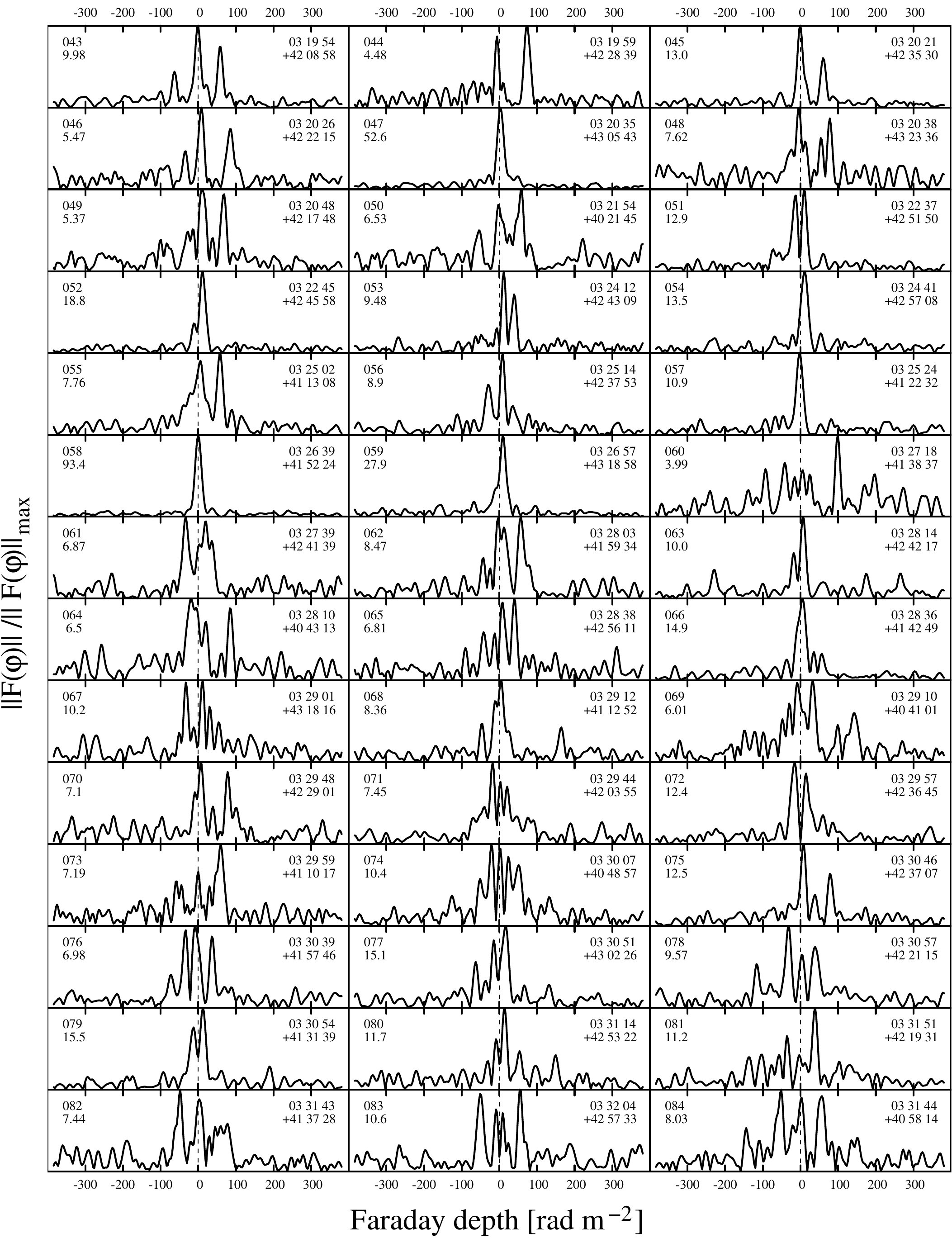}
 \end{minipage} \caption{Faraday dispersion spectra of interesting
   lines of sight through the mosaic. The top left number is the
   label of the spectrum in
   Fig.~\ref{brentjens_perseusmosaic_fig:imap}. The vertical scale is
   different for each graph. The peak brightness in
   mJy~beam$^{-1}$~rmsf$^{-1}$ is displayed below the number of the
   spectrum. The top right corner of each plot l
ists the J2000
   $\alpha$ and $\delta$.}
\label{brentjens_perseusmosaic_fig:spectra-1}
\end{figure*}

The size of the mosaic is $4\fdg5$ in declination by $7\degr$ in right
ascension. Figure~\ref{brentjens_perseusmosaic_fig:imap} shows a full
resolution Stokes $I$ map of the area. It is the average of 143
channels with an average frequency of 351~MHz. The peak flux is
19.97~Jy~beam$^{-1}$ at the location of \object{3C~84}. The actual RMS
noise level of the total intensity map ranges from 1.5~mJy~beam$^{-1}$
in the most western fields, A and B, to 3~mJy~beam$^{-1}$ in field G.
The difference between the expected thermal noise and the observed RMS
noise is caused by ionospheric non-isoplanaticity. It is worse
in the eastern fields because they contain more bright sources than
the western fields. These problems are unpolarized and therefore do
not limit the dynamic range of the Stokes $Q$ and $U$ images. The
total dynamic range is approximately 6000:1. The labelled crosses
indicate the positions of the Faraday dispersion spectra presented in
Fig.~\ref{brentjens_perseusmosaic_fig:spectra-1}. There is no significant
\modified{Stokes $I$ source at} most of these locations.

The individual Faraday dispersion spectra (hereafter simply spectra)
in Fig.~\ref{brentjens_perseusmosaic_fig:spectra-1} show that emission
at multiple Faraday depths along the same line of sight are the norm,
rather than the exception in this part of the sky. The Faraday depth
of significant emission ranges from $-60$~rad~m$^{-2}$ (spectrum 43)
to $+100$~rad~m$^{-2}$ (spectrum 60), or perhaps $+140$~rad~m$^{-2}$
if the complexity in spectrum 69 is real. Spectra 1, 6, and 37 show
the quality of the RM-CLEAN: all have residuals of less than 2\% of
the main peak, which is comparable to the RMS noise of
\modified{nearby} pixels.

Spectra 24, 28, and 31 are lines of sight through the ``lens''
structure. They all show clear peaks at a Faraday depth of
approximately $+50$~rad~m$^{-2}$, in addition to peaks near
$+6$~rad~m$^{-2}$.  Spectrum 30 goes through the centre of the
``doughnut'' and is triple valued. Spectrum 43 goes straight through the
``blob'' directly north of the extended tail of \object{NGC~1265}. The
most complex spectra are located in the fields G and H. The highest
absolute Faraday depths occur in fields E, G, and H.

Figures~\ref{brentjens_perseusmosaic_fig:rmcube-1} through
\ref{brentjens_perseusmosaic_fig:rmcube-4} show the most interesting
part of the RM-cube. The first few frames are devoid of significant
emission. The arc between $-72$ and $-60$~rad~m$^{-2}$ is instrumental
and is caused by a minor calibration error of unknown origin in Stokes
$U$ of field G. The first significant emission appears at a Faraday
depth of $-48$~rad~m$^{-2}$ in the northern four fields, especially in
fields A (north-west) and H (north-east). The patches have structure
at scales of a few arc minutes. The emission increases particularly in
the north-eastern part of the mosaic when the Faraday depth approaches
0~rad~m$^{-2}$.

The entire mosaic is filled with emission with structure at typical
scales of tens of arc minutes at Faraday depths between $-6$ and
$+12$~rad~m$^{-2}$. The peak brightness is almost
30~mJy~beam$^{-1}$~rmsf$^{-1}$ in the north-west corner. This type of
emission dissolves at $\phi\approx +18$~rad~m$^{-2}$. At that point, a
well-defined linear structure develops between
$\alpha\approx3^\mathrm{h}12^\mathrm{m}$, $\delta\approx+39\fdg7$ and
the north-west corner of the mosaic. The following frames show that
the emission slowly moves east with increasing Faraday depth. It also
becomes less uniform. The thin straight line that runs from
$\alpha=3^\mathrm{h}16^\mathrm{m}$, $\delta\approx +40\degr$ to
$\alpha=3^\mathrm{h}6^\mathrm{m}$, $\delta\approx +43\fdg5$ at
$\phi=+42$~rad~m$^{-2}$ is called the ``front'' in
\citet{DeBruynBrentjens2005}. There are several highly significant
structures in the area between $\alpha\approx
3^\mathrm{h}8^\mathrm{m}$~--~$\alpha\approx 3^\mathrm{h}8^\mathrm{m}$,
$\delta\approx +41\fdg5$ and the northern edge of the mosaic. There
are also small patches of emission across the rest of the map,
particularly in the north-eastern area.

The ``doughnut'' and brightest parts of the ``lens''
\citep{DeBruynBrentjens2005} are visible at Faraday depths of $+48$
and $+54$~rad~m$^{-2}$, along with several patches north of them.
There is a blob of emission at $\phi=+60$~rad~m$^{-2}$ around line of
sight 43, north of \object{3C~84} and directly north of the extended
tail of \object{NGC~1265}. The ``bar'' \citep{DeBruynBrentjens2005} is
visible around line of sight 44 at a Faraday depth of
$+78$~rad~m$^{-2}$. There are still several significant patches of
polarized emission in fields E and H, which fade away at Faraday
depths above 100~rad~m$^{-2}$.



\section{Discussion}
\label{brentjens_perseusmosaic_sec:discussion}


\modified{The bright emission that spans the entire mosaic at a relatively
uniform Faraday depth of 0 to +12~rad~m$^{-2}$ is evidently Galactic}:
its spatial structure, Faraday depth, and brightness temperature are
typical for medium latitudes and comparable resolutions
\citep{UyanikerEtAl1999,HaverkornKatgertDeBruyn2003,
  HaverkornKatgertDeBruyn2003c,SchnitzelerEtAl2007}. The brightness
temperature of the polarized \modified{intensity is} 5 to 10~K with a maximum of
14~K\footnote{\modified{At 351~MHz, 10~mJy~beam$^{-1}$ of $2\arcmin\times3\arcmin$ FWHM
corresponds to a brightness temperature of 4.6~K}}. The Faraday depth
range is consistent \modified{with observations} by
\citet{HaverkornKatgertDeBruyn2003} at similar $l$ and $|b|$.

In the remainder of this section \modified{I argue} that most, if not all, of the
other extended polarized emission at both higher and lower Faraday
depths \modified{is Galactic} and is not associated with the Perseus cluster. I
will do that by discussing the arguments mentioned in the introduction
in the light of the new observations.

\subsection{A special Faraday depth}

In Fig.~\ref{brentjens_perseusmosaic_fig:spectra-1}, spectra 24, 28,
and 31 clearly show the separation between the emission at low $\phi$
and the ``front'' and the ``doughnut''. It is also evident in spectrum 43
(the ``blob'') and spectra 44, 46, and 49 (the ``bar''). However, as can
be seen in the images in
Figs.~\ref{brentjens_perseusmosaic_fig:rmcube-1} to
\ref{brentjens_perseusmosaic_fig:rmcube-4} and the Faraday dispersion
spectra in Fig.~\ref{brentjens_perseusmosaic_fig:spectra-1}, there is
significant emission at \emph{all} Faraday depths between
$-48$~rad~m$^{-2}$ and $+100$~rad~m$^{-2}$.

The strong emission at $\phi=+6$~rad~m$^{-2}$ \modified{connects smoothly to} the
emission at higher Faraday depth at $\phi=+18$~rad~m$^{-2}$ in fields
A and B. Spectrum 55 contains emission at $\phi =
+60$~rad~m$^{-2}$. Several spectra to its east contain \modified{emission
between} $+50$~rad~m$^{-2}$ and $+100$~rad~m$^{-2}$ (spectrum 60). See
for example the patches between lines of sight 60, 82, 75, and 61 at
$\phi=+84$~rad~m$^{-2}$.

\modified{When considering the entire mosaic, there is no trend of
  higher Faraday depths closer to \object{3C~84}.}  Of course there
is \modified{a west} to east gradient between $\phi=+18$~rad~m$^{-2}$ and
$\phi=+60$~rad~m$^{-2}$, \modified{but higher} and lower Faraday depths
occur throughout the mosaic. The highest absolute Faraday depths \modified{and
most} complex Faraday dispersion spectra occur in fields E, G, and
H in areas that can not be associated with the \object{Perseus~cluster}.

\modified{The Faraday depths} of the ``front'' ($+42$ to $+48$~rad~m$^{-2}$),
``lens'' ($\approx +50$~rad~m$^{-2}$), ``doughnut'' ($\approx
+50$~rad~m$^{-2}$), ``blob'' ($+60$~rad~m$^{-2}$), and ``bar''
($+78$~rad~m$^{-2}$) are neither extreme, nor special when compared to
the range of Faraday depths observed in this mosaic.  \modified{One can
therefore not} distinguish between Galactic polarized emission and
cluster related polarized emission in this field based solely on the
value of the Faraday depth if it is in the range from
$-48$~rad~m$^{-2}$ to $+100$~rad~m$^{-2}$.

\begin{figure} \centering
  \resizebox{\hsize}{!}{\includegraphics{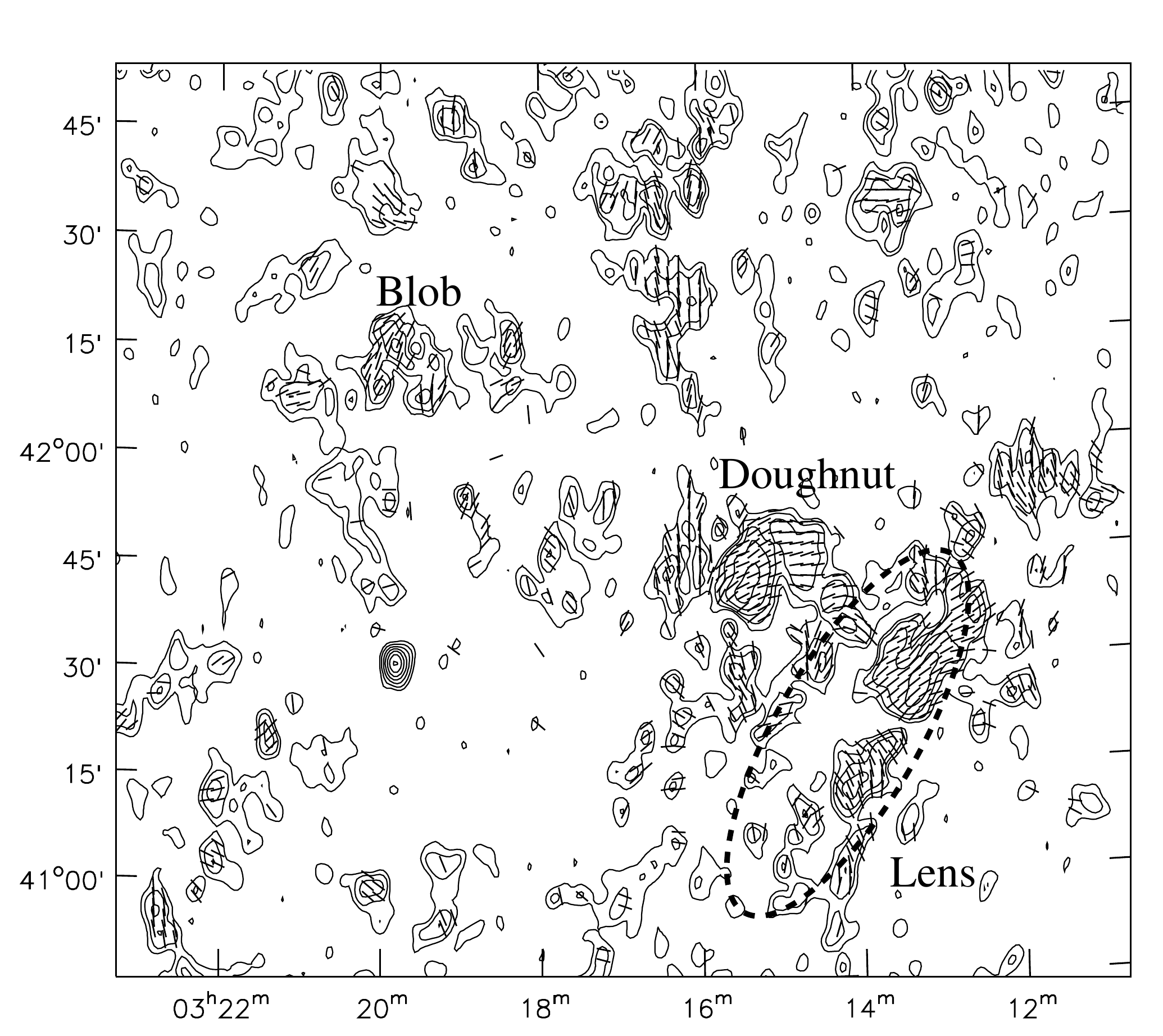}}
  \caption{Polarization vectors at 350.22~MHz overlaid on contours of
    $|F(\phi)|$ at $\phi = +51$~rad~m$^{-2}$ of the ``doughnut'' area. The contours start at
    2.5~mJy~beam$^{-1}$~rmsf$^{-1}$ and are scaled by a factor of
    $\sqrt{2}$.}
\label{brentjens_perseusmosaic_fig:lens-doughnut-zoom}
\end{figure}

\subsection{Smaller spatial scales at high $|\phi|$}

\begin{figure}
\centering
\resizebox{\hsize}{!}{\includegraphics{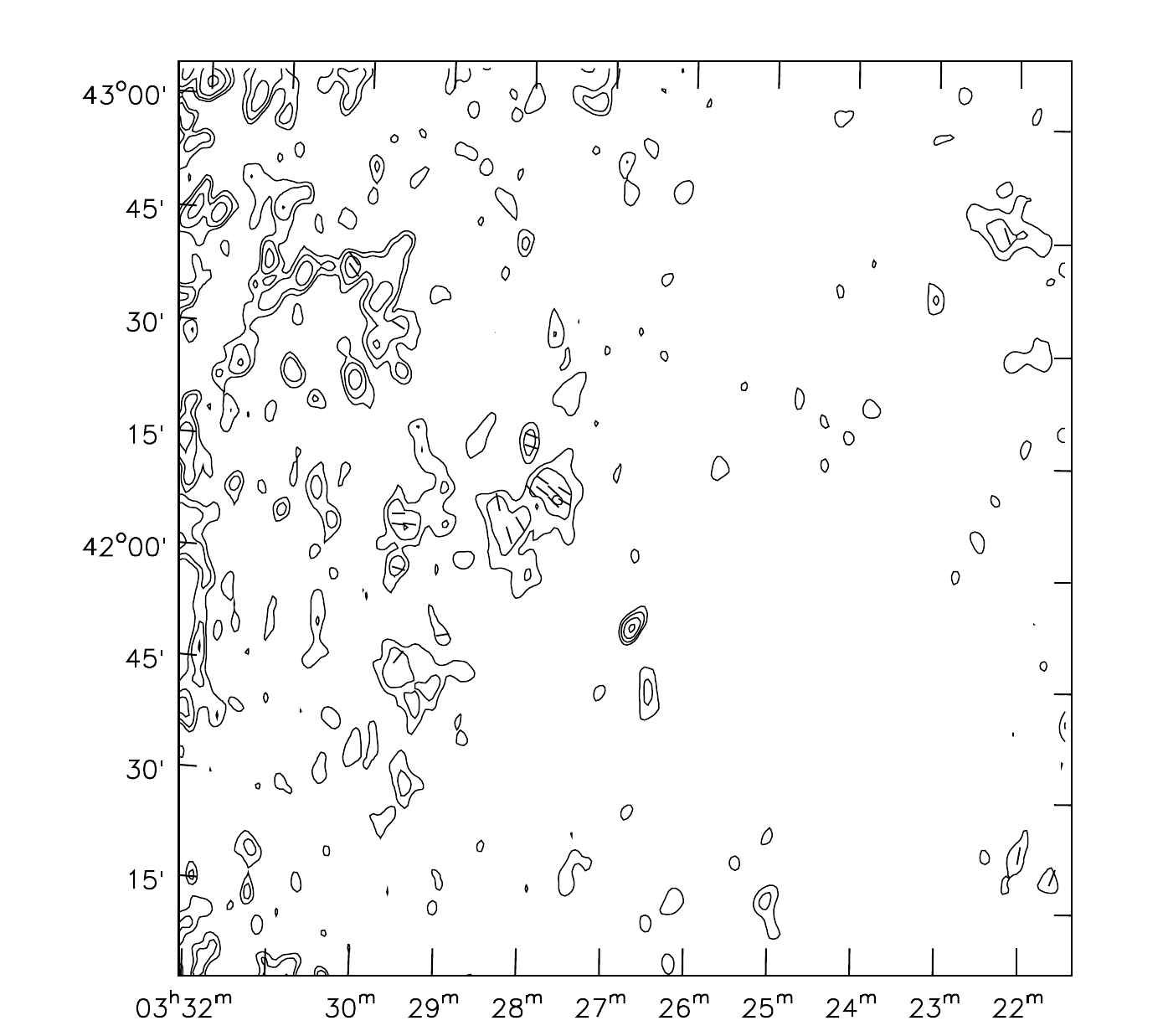}}
\caption{Polarization vectors at 350.22~MHz overlaid on contours of
$|F(\phi)|$ at $\phi = +84$~rad~m$^{-2}$ of the intersection of
fields G and H.  The contours start at
2.5~mJy~beam$^{-1}$~rmsf$^{-1}$ and are scaled by a factor of~$\sqrt{2}$.}
\label{brentjens_perseusmosaic_fig:patches+84}
\end{figure}

Figures~\ref{brentjens_perseusmosaic_fig:lens-doughnut-zoom}
and~\ref{brentjens_perseusmosaic_fig:patches+84} show the position
angle of the electric field on top of polarized intensity contours.
The position angles have \emph{not} been derotated to 0 wavelength
because the $1\sigma$ uncertainty of the Faraday depth at $\phi >
+18$~rad~m$^{-2}$ was generally worse than 0.5~rad~m$^{-2}$. Instead,
the vectors reflect the position angles at $\lambda_0^2 =
0.73275$~m$^{-2}$ (350.22~MHz), hence no conclusions can be drawn from
the absolute position angles in the images. The images are only used
to estimate typical scales at which the position angle changes by a
radian or more at 350.22~MHz.

The area of the ``lens'', ``doughnut'', and ``blob'' \modified{at $\phi =
+51$~rad~m$^{-2}$} is shown in
Fig.~\ref{brentjens_perseusmosaic_fig:lens-doughnut-zoom}. The lens is
difficult to recognize due to the lower signal to noise ratio compared
to the observations by \citet{DeBruynBrentjens2005}. \modified{The position}
angles are fairly uniform in patches of the order of $15\arcmin$
across, changing abruptly at the borders between these patches. The
polarized patches at the same Faraday depth in field D, north of the
``lens'', ``doughnut'', and ``blob'', are comparable, but are too far
away from \object{3C~84} to be associated with the Perseus
cluster. The polarized emission in fields G and H at
$\phi=+84$~rad~m$^{-2}$ has fairly uniform polarization angles across
each emission patch. \modified{These patches are $3\arcmin$ to
$10\arcmin\times20\arcmin$ large} (see
Fig.~\ref{brentjens_perseusmosaic_fig:patches+84}).

The polarization angle structure of a few representative images from
the full RM-cube is shown in
Figs.~\ref{brentjens_perseusmosaic_fig:polangles-1} and
\ref{brentjens_perseusmosaic_fig:polangles-2}. At $\phi =
+6$~rad~m$^{-2}$ the position angles are fairly uniform at scales of
$30\arcmin$ to $90\arcmin$. At $\phi =+30$~rad~m$^{-2}$ it changes at
$20\arcmin$ to $30\arcmin$ scales. At $\phi=+42$~rad~m$^{-2}$ the
typical scale is $10\arcmin$ to $30\arcmin$, and at higher Faraday
depths scales range from $3\arcmin$ to $20\arcmin$. These changes can
be due to differences in intrinsic polarization, changes in Faraday
rotation, or a combination of the two effects. Because of the
uncertainty of the precise Faraday depth, it is not possible to
discriminate between these possibilities.

\modified{The scale} size at which the polarization angles change \emph{does}
decrease with increasing $|\phi|$, but \modified{this is} not limited to the area
\modified{near} \object{3C~84} and \modified{is therefore no argument in favour of nor
against cluster association. Furthermore, }the scales at which the
polarization angles at 350.22~MHz change in the ``lens'',
``doughnut'', ``front'', and ``blob'' are comparable to other
structures at similar Faraday depth that can not be \modified{linked to} the
Perseus cluster.

\subsection{Fractional polarization}

The fractional polarization at 351~MHz was estimated by dividing the
polarized intensity, integrated over all Faraday depths, by the
408~MHz \citet{HaslamEtAl1982} total intensity map converted to
351~MHz using a Galactic synchrotron brightness temperature spectral
index $\beta=-2.8$
\citep{ReichReich1988A,ReichReich1988B,PlataniaEtAl1998}. Between
10~MHz and 100~MHz, the spectral index is $-2.55$ \citep{Cane1979},
hence the actual spectral index between 408~MHz and 351~MHz is
probably closer to $-2.7$. The difference with $-2.8$ is negligible
for the small extrapolation from 408~MHz to 351~MHz.

\modified{The noise in the derotated $Q$ and $U$ maps is Gaussian}:
\begin{equation}
\mathcal{P}(n)\ \mathrm{d}n = \frac {1}{\sigma\sqrt{2 \pi}}
\mathrm{e}^{-\frac{(n-\mu)^2}{2\sigma^2}}\mathrm{d}n,
\end{equation}
where $\mathcal{P}(n)\ \mathrm{d}n$ is the probability of finding a
noise value between $n$ and $n+\mathrm{d}n$, and $\mu$ and $\sigma$
are the mean and \modified{standard deviation.}  If the $Q$ and $U$ noise
distributions have equal $\sigma$, zero mean, and \modified{are uncorrelated},
then the probability of finding a \modified{noise} value of $|F|$ between $f$ and
$f+\mathrm{d}f$ is:
\begin{equation}
\mathcal{P}(f)\ \mathrm{d}f = \left\{
\begin{array}{ll}
\frac{f}{\sigma^2} \mathrm{e}^{-\frac{f^2}{2\sigma^2}}\ \mathrm{d}f &
\ \mbox{if}\ f\ge 0\\
0 & \ \mbox{if}\ f < 0
\end{array}\right.
\end{equation}
The RMS of $|F|$ is equal to the RMS of $Q$ and $U$, which is
$\sigma\sqrt{2}$. The mean value of the noise in $|F|$ is
\begin{equation}
\langle f \rangle = \sigma\sqrt{\frac{\pi}{2}}.
\end{equation}

The polarized surface brightness of a \modified{low S/N} line of sight,
integrated over a range of equidistant Faraday depths
$\phi_1\cdots\phi_n$, and corrected for the non-zero mean of the noise
level, is \modified{therefore}
\begin{equation}
|P| =B^{-1}\sum_{i=1}^n \left(|F(\phi_i)| -
\sigma\sqrt{\frac{\pi}{2}}\right),
\label{brentjens_perseusmosaic_eqn:integrated-pol}
\end{equation} 
where $B$ is the area under the restoring beam of the RM-CLEAN divided
by $\Delta\phi = |\phi_{i+1}- \phi_i|$. \modified{See \citet{WardleKronberg1974}
for a more general treatment of uncertainties and biases in RM work.}

\begin{figure}
\centering
\includegraphics[width=\columnwidth]{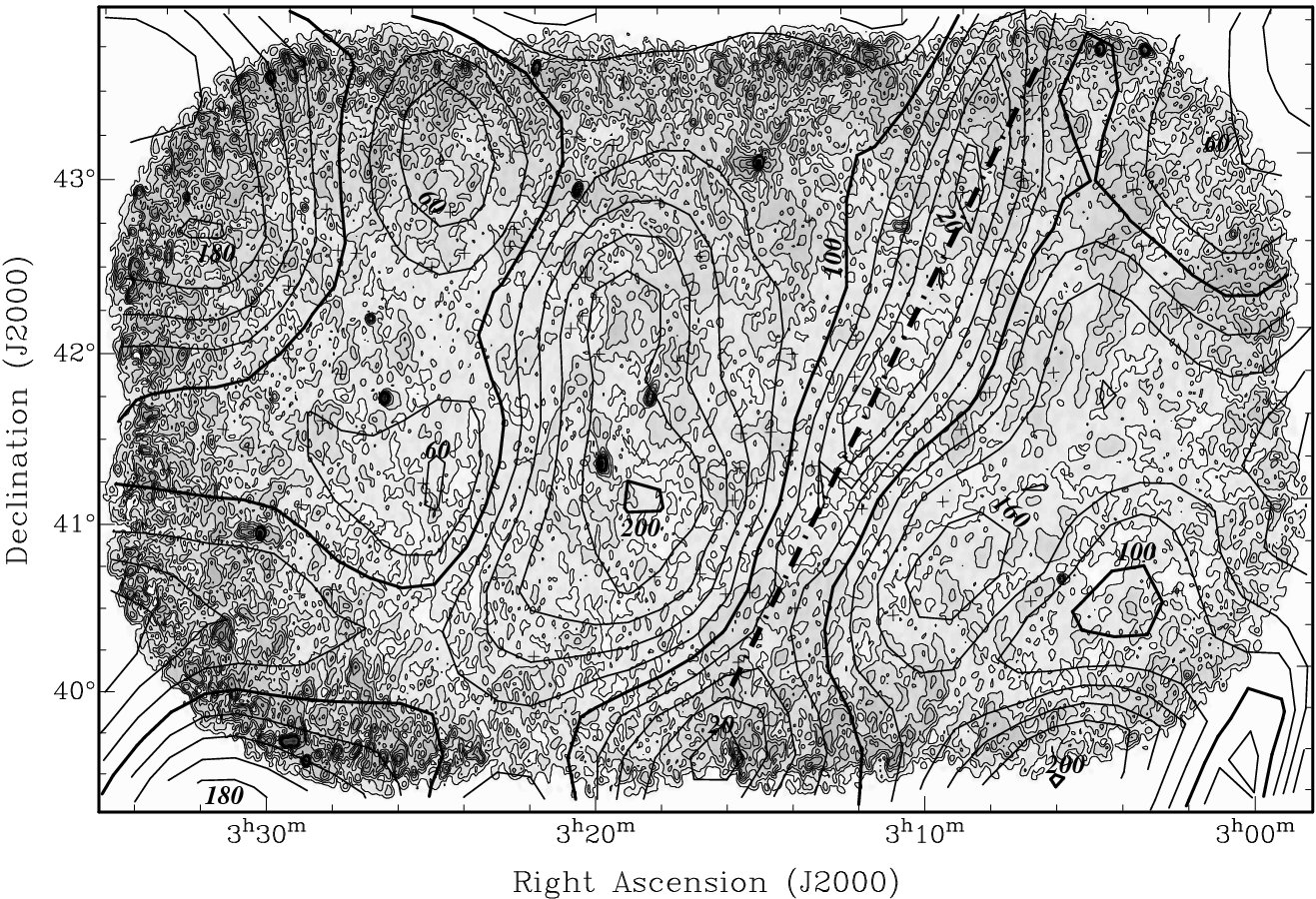}
\caption{\citet{WollebenEtAl2006} polarized intensity contours at
1.41~GHz overlaid on the integrated polarized intensity map of the RM-cube.
The thin contours represent the polarized intensity at 351~MHz and are
drawn at 4 to 40~K in steps of 4~K. 
The thick contours represent the Wolleben map and are drawn at 20 to
200~mK in steps of 20~mK. The dashed line represents the ``front''.}
\label{brentjens_perseusmosaic_fig:wolleben-overlay}
\end{figure}

\begin{figure}
\centering
\includegraphics[width=\columnwidth]{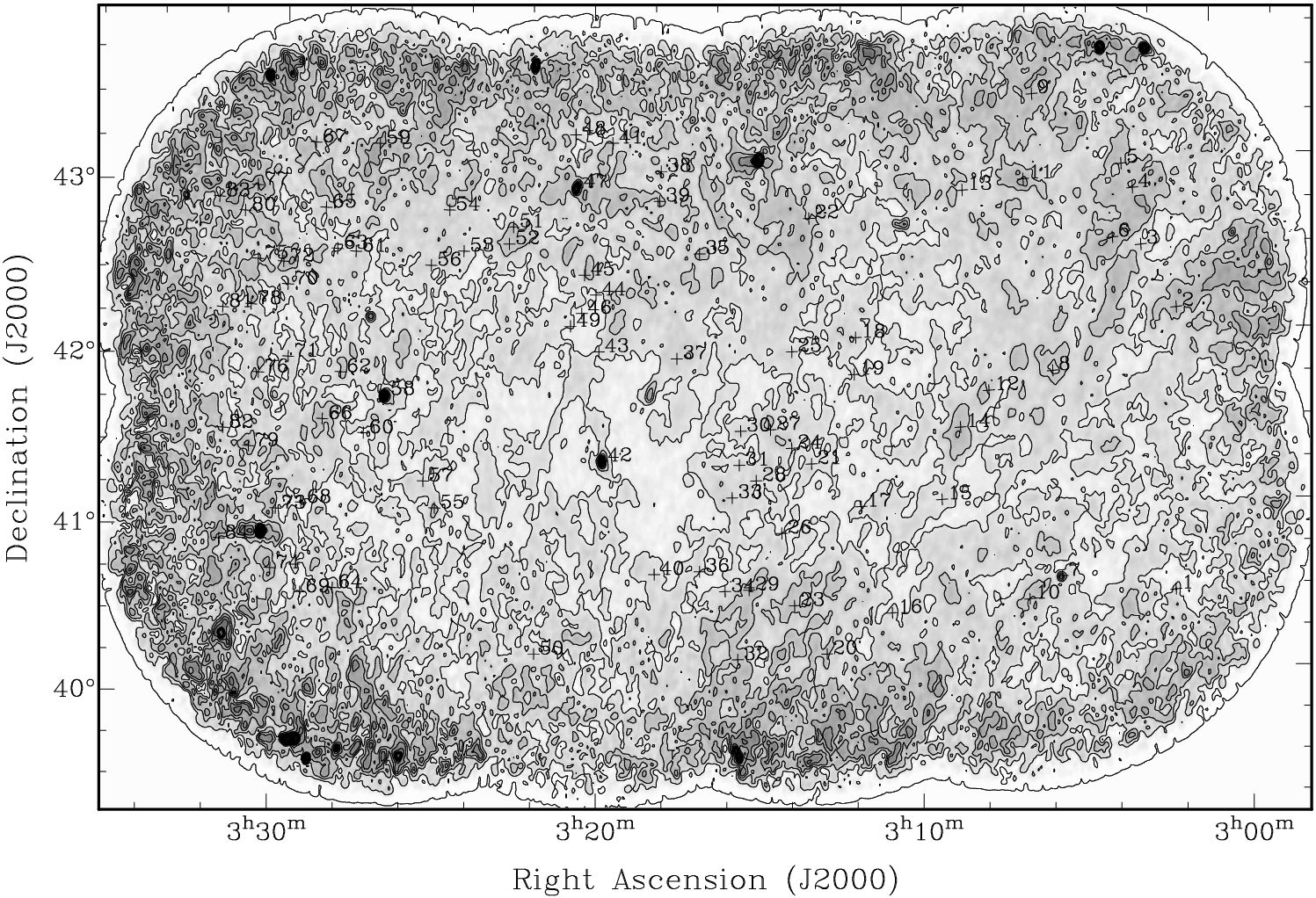}
\caption{Map of the fractional polarization at 351~MHz.  The contours
are 10\%, 20\%, 30\%, 40\%, and 50\%. The grey scale runs linearly from
0 (white) to 70\% (black).}
\label{brentjens_perseusmosaic_fig:polfraction}
\end{figure}

\citet{WollebenEtAl2006} have conducted an absolutely calibrated
survey of polarized emission north of declination $-30\degr$ at
$1.41$~GHz with the $26$~m telescope at the DRAO site \modified{at a resolution
of} $36\arcmin$\ FWHM. The integrated 351~MHz polarized intensity
overlaid with the polarized intensity contours from
\citet{WollebenEtAl2006} is shown in
Fig.~\ref{brentjens_perseusmosaic_fig:wolleben-overlay}. The noise
level in the 351~MHz map is approximately 0.5~K. With a spectral index
of $-2.8$, the brightness temperature at 351~MHz should be 50 times
higher than at 1.41~GHz. This is approximately the case in most of the
field, which implies that there is very little depolarization between
1.41~GHz and 351~MHz. In some places, the polarized intensity is even
higher at 351~MHz than one would expect based on the low resolution
polarized intensity at 1.41~GHz and a spectral index of $-2.8$.
Examples are the area containing the ``front'', ``lens'', and
``doughnut'', and the highly polarized region in field A. This is
probably caused by beam depolarization in the \citet{WollebenEtAl2006}
observations due to differences in intrinsic polarization angle at
scales well below $36\arcmin$ that are resolved in the observations
presented here.

Figure~\ref{brentjens_perseusmosaic_fig:polfraction} displays the
fractional polarization. It is mostly between 10\% and 20\% with a
maximum of 35\% in field A. Although these values are well below the
theoretical maximum of 70\%, they are relatively high. The low
fractional polarization between \object{3C~84} and \object{NGC~1265}
is an artifact caused by the low resolution ($0\fdg85$) of the
\citet{HaslamEtAl1982} map, which blends these powerful
sources. Because there are no absolutely calibrated polarimetric
single dish observations of the field \modified{near $351$~MHz}, my maps may lack
$Q$ and $U$ features at scales $\gtrsim 90\arcmin$. The \modified{fractions} are
therefore strictly speaking lower limits.

The lack of depolarization implies that the synchrotron emitting areas
have a Faraday thickness of less than 1~rad~m$^{-2}$. This \modified{is
remarkable because} the range of Faraday depths is two orders of
magnitude larger. Assuming a line of sight magnetic field of $1\ \mu$G
and a local electron density of $0.03$~cm$^{-3}$
\citep{GomezBenjaminCox2001,CordesLazio2002}, a Faraday thickness of
1~rad~m$^{-2}$ corresponds to only 40~pc, which is difficult to
reconcile with the smoothness of the Galactic synchrotron foreground
unless the structures \modified{are close} to the Sun. Assuming that the emitting
patches are approximately as thick as they are wide, \modified{non-detection of
Stokes $I$} at $90'$ scales implies that the clouds are closer than
1.6~kpc. Polarization observations at lower frequencies are required
to follow the depolarization and determine the exact \modified{Faraday
thickness.}

\begin{figure}
\centering
\includegraphics[angle=-90,width=\columnwidth]{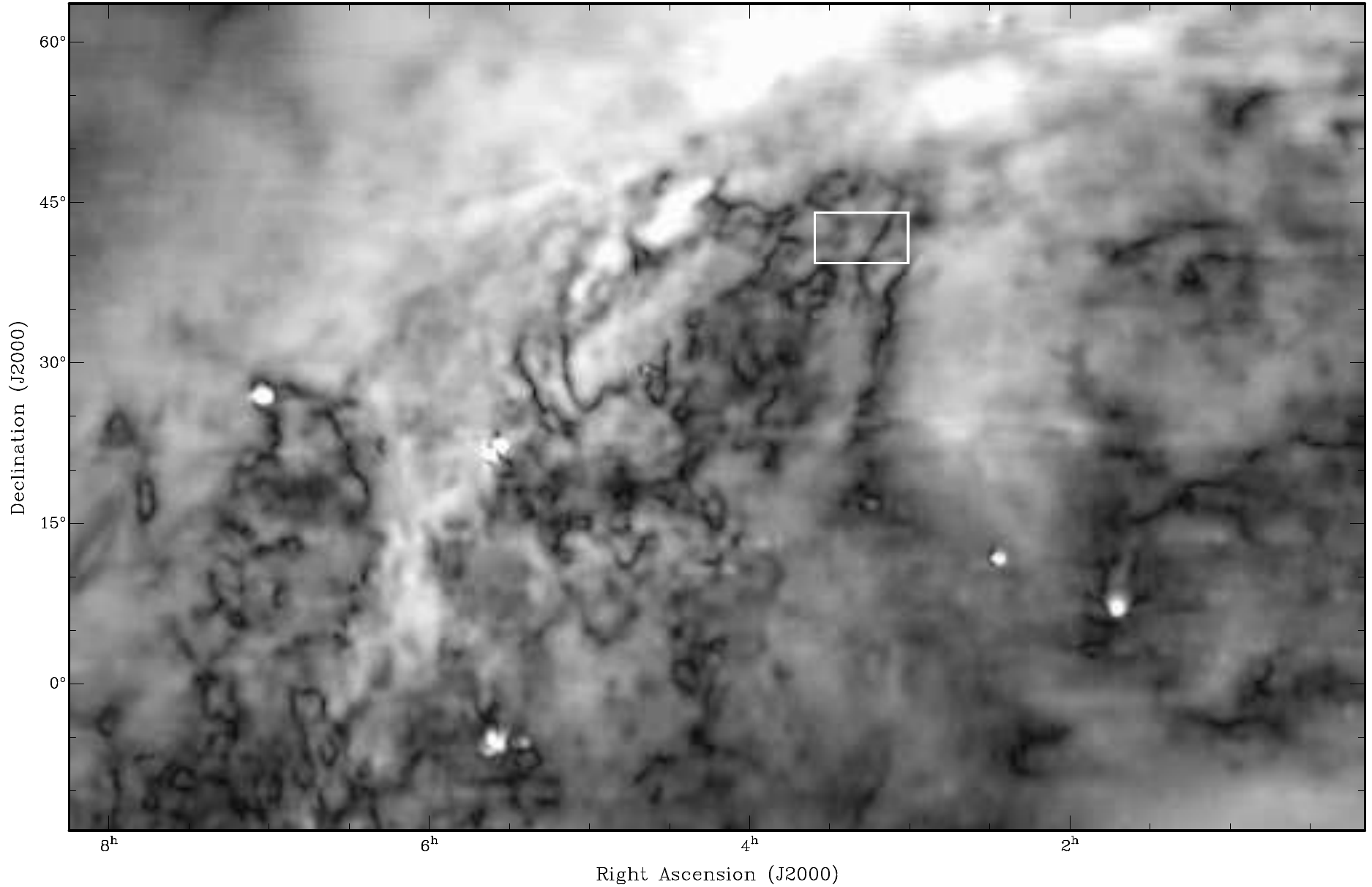}
\caption{Part of the \citet{WollebenEtAl2006} polarized intensity
map at 1.4~GHz. The position and extent of the Perseus mosaic is
indicated by the white rectangle. The grey scale is proportional to the
square root of the polarized intensity and runs from 0 (black) to
500~mK (white).}
\label{brentjens_perseusmosaic_fig:wolleben-wide}
\end{figure}

\subsection{The ``front''  and the Perseus-Pisces super cluster}

The ``front'' was tentatively interpreted by
\citet{DeBruynBrentjens2005} as a large scale structure formation
shock at the interface between the Perseus cluster and the
Perseus-Pisces super cluster. It was unclear at that time whether the
``front'' extended much beyond the primary beam of the WSRT. As can be
seen in the image at $\phi=+42$~rad~m$^{-2}$ in
Fig.~\ref{brentjens_perseusmosaic_fig:rmcube-3}, it does. The location
of the ``front'' is indicated by the dashed line in
Fig.~\ref{brentjens_perseusmosaic_fig:wolleben-overlay}. It runs from
line of sight 32 via lines of sight 18 and 12 to line of sight 19.
Association with the Perseus cluster is therefore unlikely.

\modified{As can be seen in the 21~cm polarization map by
\citet{WollebenEtAl2006}
(Fig.~\ref{brentjens_perseusmosaic_fig:wolleben-wide}), the Perseus
cluster is located behind the north-western tip of a large field of
Galactic depolarization canals (see e.g.  \citet{FletcherShukurov2006}
and \citet{HaverkornKatgertDeBruyn2000} for an in-depth treatment of
depolarization canals). Interestingly, the ``front'' coincides with
the centre of such a canal, hence the ``front'' is very likely
Galactic. If the ``lens'' is associated with the ``front'', it must
also be Galactic. The two structures may of course be unrelated, but
their coincidence in Faraday depth, position, and position angle
suggests otherwise.}

\subsection{$\phi$ with respect to background sources}

\begin{figure}
\centering
\includegraphics[width=\columnwidth]{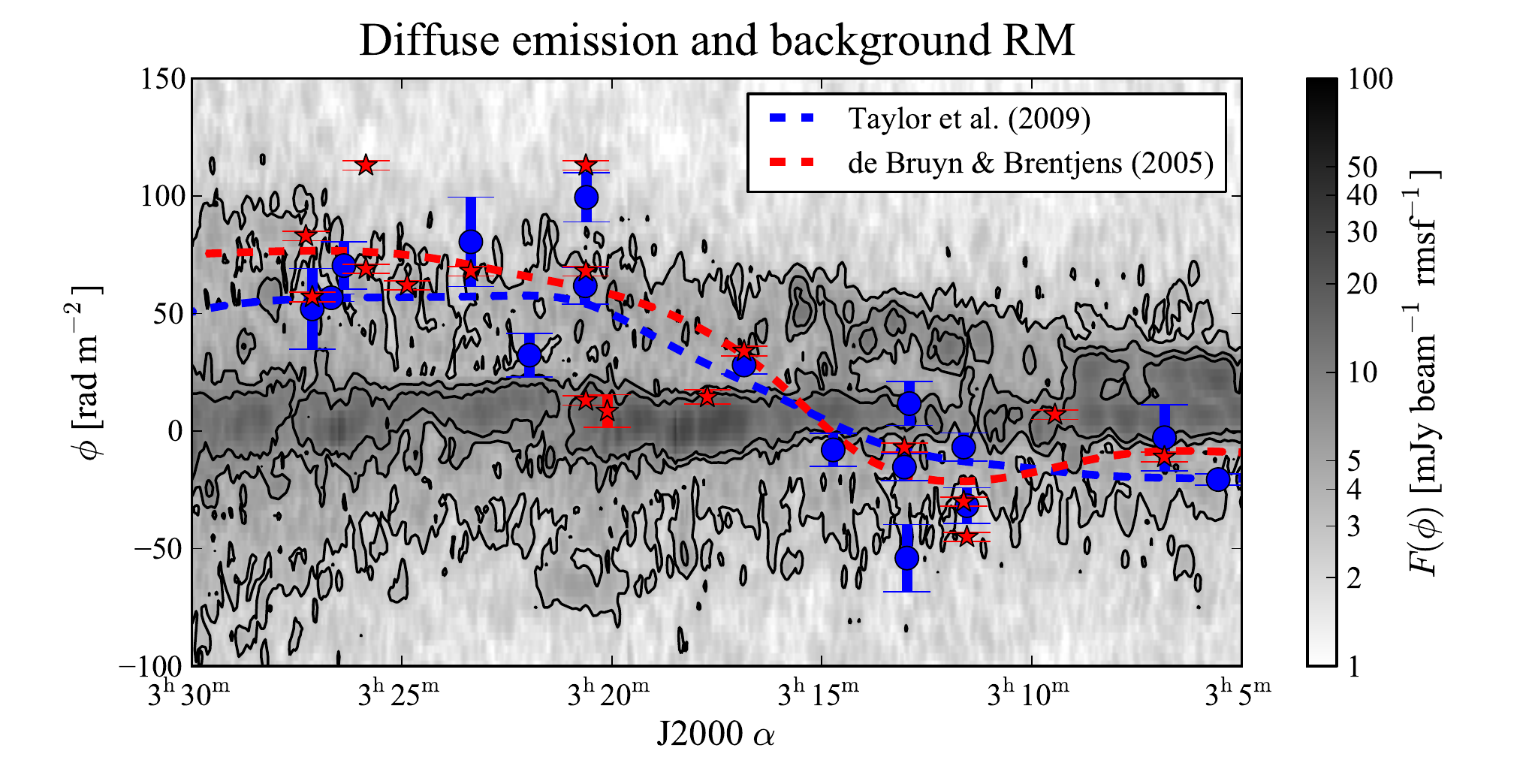}
\caption{Diffuse emission with respect to background RM. The grey
  scale image is the logarithm of the maximum brightness in a 1 degree
  wide slab. The background sources are from a 4 degree wide
  slab. Both slabs are centred at $\delta= 41\degr 48\arcmin
  12\arcsec$. Contour levels are at 3, 6, and
  9~mJy~beam$^{-1}$~rmsf$^{-1}$.}
\label{brentjens_perseusmosaic_fig:comparison-with-background}
\end{figure}

The excess of $+40$ to $+50$~rad~m$^{-2}$ in $\phi$ of the structures
observed by \citet{DeBruynBrentjens2005} with respect to the
background sources was based on a small number of polarized sources
near the centre of the mosaic. \citet{TaylorStilSunstrum2009} have
since published a comprehensive RM catalogue based on a re-analysis of
37\,543 NVSS sources, allowing a more detailed analysis. de~Bruyn et
al. (in prep.) have \modified{also} conducted WSRT observations of more than 200
polarized sources in and around this area during the 2004/2005 winter
season. Those data will be reported in a subsequent paper.

Figure~\ref{brentjens_perseusmosaic_fig:comparison-with-background}
illustrates the relation between the \citet{TaylorStilSunstrum2009}
\modified{sources, the \citet{DeBruynBrentjens2005} sources, and the diffuse
polarized emission}. The logarithmic grey scale image represents the
\emph{maximum} $|F(\phi)|$ at a particular Faraday depth and
horizontal position in a $1\degr$ thick horizontal slab through the
RM-cube\modified{, centred at} $\alpha =3^\mathrm{h}20^\mathrm{m}$ and $\delta=
+41\degr 48\arcmin 12\arcsec$. The \citet{TaylorStilSunstrum2009}
sources are those within a $4\degr$ thick horizontal slab centred at
the same \modified{position}. \modified{The sources} from \citet{DeBruynBrentjens2005}
partly overlap with the \citet{TaylorStilSunstrum2009} selection.
Where they do, the rotation measures agree within the error bars. The
\modified{$\alpha$} tick marks indicate the right ascension at the centre of the
slabs. The dashed lines are the background points convolved with a
Gaussian kernel of $\Delta\alpha=5^\mathrm{m}$ FWHM.

Both dashed curves show a clear trend in the background RM. \modified{The same
trend is visible} in the polarized emission at $\phi > +12$ and $\alpha
< 3^\mathrm{h}17^\mathrm{m}$.  This suggests that there is an area
behind the emission at high $\phi$ with a relatively uniform Faraday
thickness of $-45\pm5$\ rad~m$^{-2}$. The fact that the excess extends
to the westernmost edge of the image rules out a cluster origin.

The scatter in background RMs is relatively large, as is the scatter
in the Faraday depth of polarized emission. Whether this scatter can
be explained adequately by a turbulent magnetized ISM or by the IGM
around the background sources needs to be investigated using numerical
MHD simulations.

\subsection{A crude model}

\begin{figure}
  \resizebox{\hsize}{!}{\includegraphics{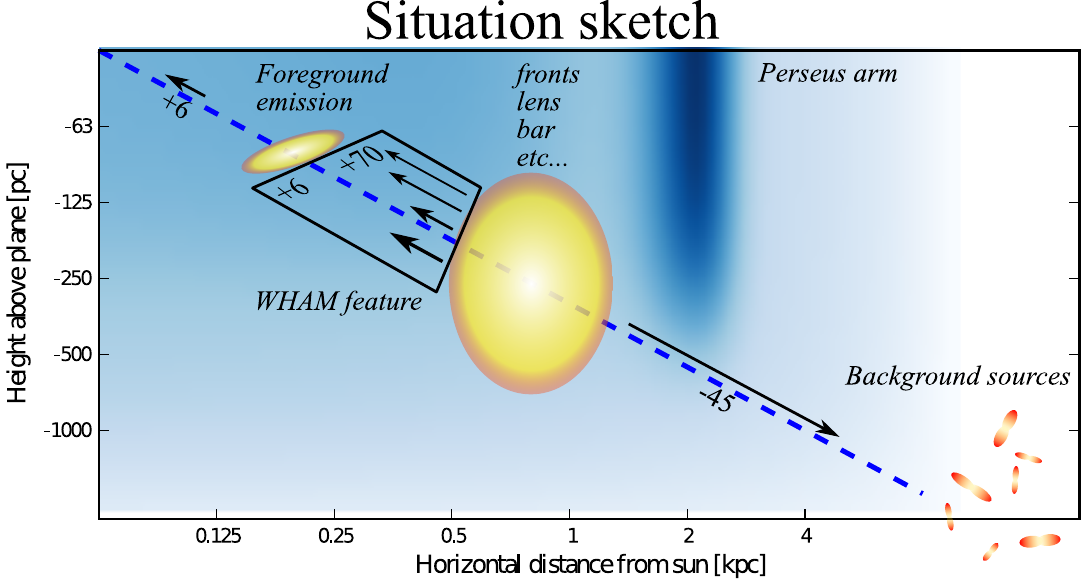}}
  \caption{Sketch of the possible layout of emission and magnetic
    fields in the observed field. Solid arrows indicate orientation
    and strength of $B_\parallel$. Yellow indicates synchrotron
    emission and blue the thermal electron density. The image is a
    vertical slice through the southern half of the Milky Way at $l =
    150\degr$. The dashed line is the line of sight towards the
    Perseus cluster at $b = -13\degr$.}
\label{brentjens_perseusmosaic_fig:galactic-model}
\end{figure}

The large scale magnetic field in the vicinity of the Sun is estimated
at 1.4~$\mu$G and points towards $l \approx 80\degr$
\citep{HanQiao1994,SunEtAl2008}.  The Faraday depth at $l \approx
150\degr$ should therefore be slightly negative, which is clearly not
observed.  The most prominent polarized features are instead observed
at positive Faraday depth, indicating a magnetic anomaly in the
direction of the \object{Perseus cluster}.

It is interesting to briefly explore the conditions in the ISM that
are required to explain the observed Faraday depths.
Figure~\ref{brentjens_perseusmosaic_fig:galactic-model} illustrates
the emerging picture. Because of the generally high degree of
polarization and lack of depolarization between L band and 350~MHz, it
is likely that the areas with the most pronounced polarized emission
are less than 40~pc thick.

The wide spread emission at $\phi\approx +6$~rad~m$^{-2}$ is probably
the most nearby component because of its uniform Faraday depth and
polarization angle structure at large spatial scales. The uniform
Faraday depth also suggests that $n_\mathrm{e}B_\parallel$ towards the
Perseus cluster near the Sun is rather uniform. The thickness of this
Faraday rotating layer is therefore
\begin{equation}
  d_\mathrm{+6} \approx 250
 \left[\frac{0.03\ \mathrm{cm}^{-3} \times 1\ \mu\mathrm{G}}{\langle
     n_\mathrm{e}|B_\parallel|\rangle}\right] \ \mathrm{pc}.
\end{equation}
If the polarized emission at $\phi=+6$~rad~m$^{-2}$ is located closely
behind this area, this would place the emission near the edge of the
local bubble, which is estimated to have a radius of approximately
200~pc \citep[see e.g.][]{SunEtAl2008}. This is consistent with the
suggestion by \citet{WollebenEtAl2006} that the Sun resides inside a
synchrotron emitting region, \modified{provided that the emission lies beyond a
non-emitting magnetized plasma.}

Beyond this emission follows an area with a Faraday thickness of
approximately $+6$ in the west to $+70$~rad~m$^{-2}$ at the centre of
the mosaic, as was discussed in the previous section. This layer is
followed by the emission containing the western diagonal structures
and the ``front'', ``lens'', ``doughnut'', and ``bar'' near the centre
of the mosaic. The diagonal features in this area -- including the
``front'' and ``lens'' -- appear connected spatially as well as in
Faraday depth.

\begin{figure}
\centering
\includegraphics[width=\columnwidth]{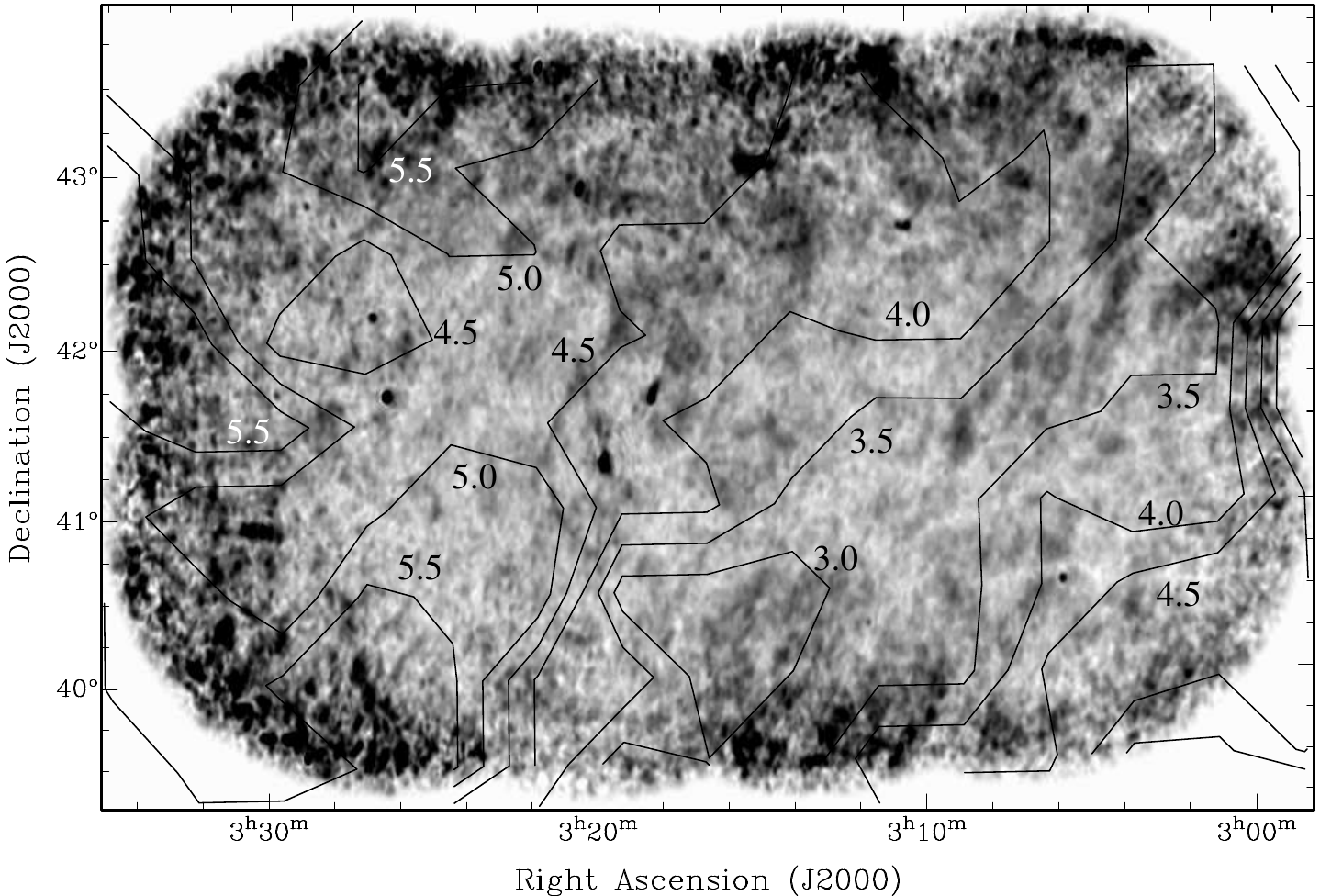}
\caption{Integrated polarized intensity at 351~MHz overlaid with WHAM H$\alpha$
surface brightness contours in Rayleighs. The grey scale runs from 0 (white) to
40~mJy~beam$^{-1}$ (black). The resolution of the polarized intensity
map is $2\arcmin\times3\arcmin$ FWHM.}
\label{brentjens_perseusmosaic_fig:wham-overlay}
\end{figure}

Figure~\ref{brentjens_perseusmosaic_fig:wham-overlay} shows the
integrated H$\alpha$ surface brightness contours observed by the
Wisconsin H$\alpha$ Mapper \citep[WHAM,][]{HaffnerEtAl2003} overlayed
on the integrated polarized intensity map. There is an absorption
feature in the integrated H$\alpha$ map running from the centre of the
southern edge of the field to the north-west corner. The minimum
H$\alpha$ brightness coincides with the bright, complex polarized
patch around line of sight 32 that is \modified{visible between $\phi =
-24$~rad~m$^{-2}$ and $+36$~rad~m$^{-2}$. Furthermore, the 60~$\mu$m
and 100~$\mu$m IRAS \citep{NeugebauerEtAl1984} infrared maps show
enhanced infrared emission here, as well as diagonally towards the
north west.} The centre of the dark H$\alpha$ feature runs along the
diagonal structures in the western part of the field between
$\phi=+24$~rad~m$^{-2}$ and $\phi=+36$~rad~m$^{-2}$. The WHAM feature
is visible between $-20$ and $+20$~km~s$^{-1}$, but is most prominent
between $-20$ and $0$~km~s$^{-1}$. Assuming a central velocity of
$-5\pm5$~km~s$^{-1}$ and a Galactic rotational velocity near the Sun
of $200\pm10$~km~s$^{-1}$ \citep{Merrifield1992,BinneyDehnen1997}, the
kinematic distance to the cloud is $0.5\pm0.5$~kpc. The polarized
emission at high Faraday depth is probably located behind the WHAM
structure.

\begin{figure}
\centering
\includegraphics[width=\columnwidth]{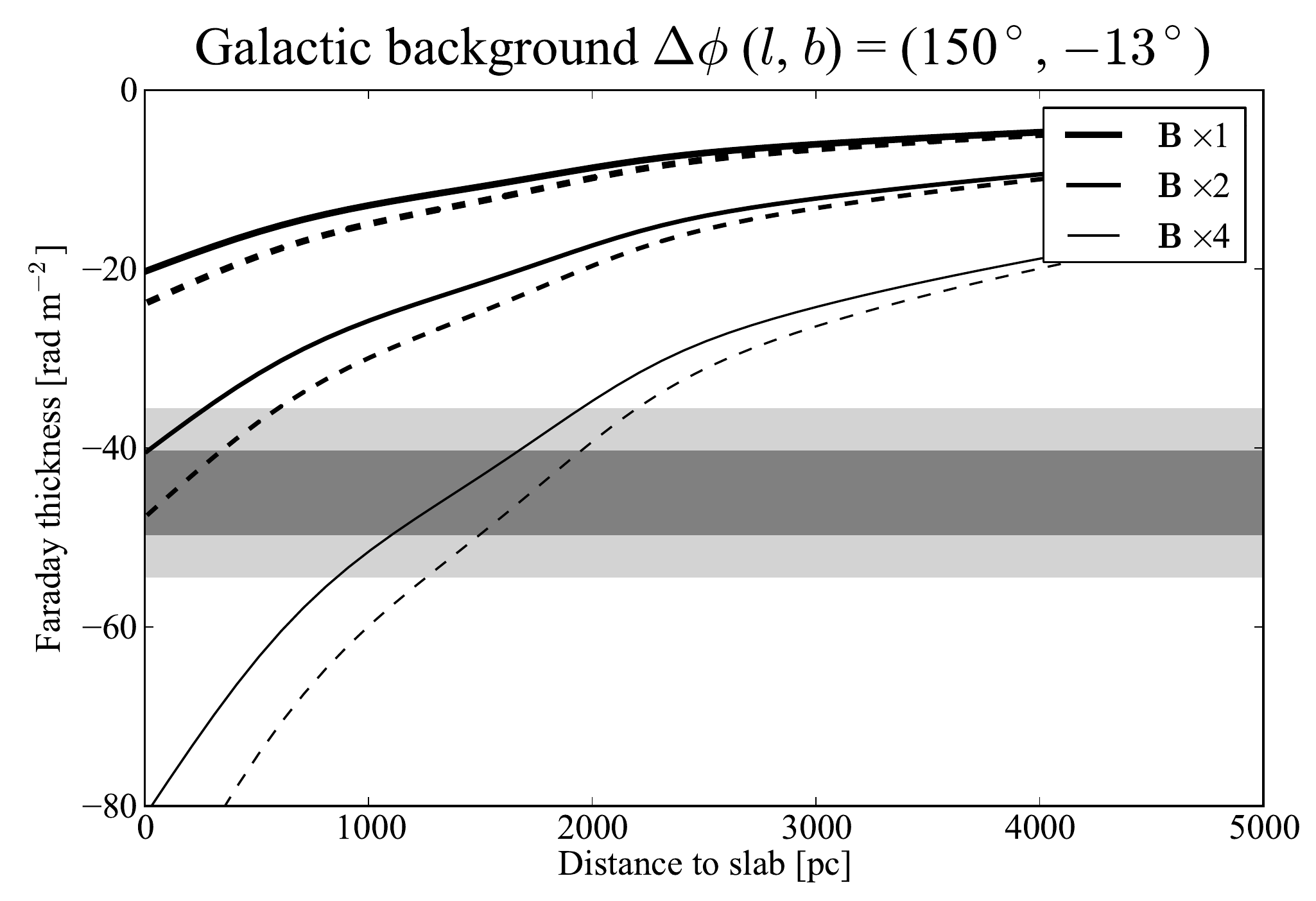}
\caption{Model Faraday thickness of a slab along the line of sight
  towards the Perseus cluster as a function of the distance to the
  near side of the slab. The far side of the slab is held constant at
  20~kpc distance from the Sun. The assumed magnetic pitch angle is
  $12\degr$ (solid lines) or $8\degr$ (dashed lines). The thinner sets
  of lines represent magnetic models where the total field strength
  was multiplied with a factor of two or four. The grey bar indicates
  the $1\sigma$ and $2\sigma$ uncertainty levels of the observed Faraday
  thickness of the slab between the high-$\phi$ polarized emission and
  the polarized background sources.}
\label{brentjens_perseus_mosaic_fig:halo-rm}
\end{figure}

Another hint at the proximity of the high-$\phi$ emission comes from
the $-45\pm5$~rad~m$^{-2}$ offset in Faraday depth between the
high-$\phi$ emission and the polarized background sources. Assuming
that most of this offset is due to the Milky Way, one can estimate the
Faraday thickness of this layer using models for the electron density
and magnetic field along the line of sight, and a lower and upper
integration limit. The electron model consists of the NE2001 spiral
arms, thin disc, and thick disc \citep{CordesLazio2002} modified
according to \citet{GaenslerEtAl2008}. The magnetic field model is the
ASS+ring model from \citet{SunEtAl2008}. The upper integration limit
was set at 20~kpc from the Sun. Simulations were performed with
magnetic pitch angles of $12\degr$ \citep{SunEtAl2008} and $8\degr$
\citep{HanQiao1994} and with different total field strengths (once,
twice, and four times the strength from \citet{SunEtAl2008}). The
results are shown in Fig.~\ref{brentjens_perseus_mosaic_fig:halo-rm}.

The points where the curves intersect with
$\Delta\phi=-45$~rad~m$^{-2}$ mark the maximum distance to the slab
-- and therefore to the most remote polarized emission -- that can still
build up the required Faraday depth when integrating out to 20~kpc. It
is clear that the default models cannot explain the observed gap. It
is necessary to either invoke a special area just beyond the most
distant emission with strongly deviating electron densities and/or
magnetic fields, or an increase in the large scale field strength or
electron density, possibly combined with a decrease in pitch angle.
In any case it is difficult to defend a distance of more than 1~kpc to
the near \modified{side} of the Faraday rotating area, hence it is likely that
the high-$\phi$ polarized emission is located well within 1~kpc from
the Sun. I consider the order of the structures in
Fig.~\ref{brentjens_perseusmosaic_fig:galactic-model} accurate. The
uncertainties in the distances to individual objects are of order a
factor of two for each of the objects.

%


\section{Concluding remarks}
\label{brentjens_perseusmosaic_sec:conclusions}

I have shown that the polarized Galactic radio synchrotron foreground
near $l\approx 150\degr$, $b\approx -13\degr$ is very complex. Most
lines of sight show radio emitting screens at multiple Faraday depths
between $-50$ and $+100$~rad~m$^{-2}$. Because of the layer of
negative Faraday depth behind the high-$\phi$ emission in this part of
the sky, it is very difficult to distinguish between Galactic and
cluster related polarized emission.


Although the ``lens'' could very well be associated with the
``front'', it remains a peculiar structure. If the ``lens'' is related
to the Perseus cluster, it is not unlike the giant curved relic
sources in \object{Abell~3667} \citep{RottgeringEtAl1997} and
\object{Abell~2744} \citep{OrruEtAl2007}. However, it is uncertain how
\modified{many are} still highly polarized at 350~MHz. \modified{Although the
\object{Abell~2256} relic is highly polarized at 1.4~GHz
\citep{ClarkeEnsslin2006}, WSRT observations at 350~MHz
\citep{Brentjens2008} showed it is
completely depolarized due to internal Faraday dispersion
\citep{Burn1966}}. Nor was there evidence for other
polarized emission in or near \object{Abell~2256} at
350~MHz. \object{Abell~2255} is another cluster with large, shock
related radio filaments at its outskirts
\citep{GovoniEtAl2005,PizzoEtAl2010}. Although these sources are
highly polarized at 1.4~GHz, they are fully depolarized at 350~MHz and
150~MHz \citep{PizzoEtAl2010}.

The arguments presented in the previous Section and the absence of
polarized relic emission at low frequencies in two other clusters with
clear evidence for merger shocks at high frequencies leads to the
conclusion that all polarized diffuse emission described in
\citet{DeBruynBrentjens2005} and \modified{in this work is Galactic and
resides within a few hundred parsecs from the Sun.}

\begin{acknowledgements}
  The Westerbork Synthesis Radio Telescope is operated by ASTRON
  (Netherlands Foundation for Research in Astronomy) with support from
  the Netherlands Foundation for Scientific Research (NWO). The
  Wisconsin H-Alpha Mapper is funded by the National Science
  Foundation. I thank A.~G.~de~Bruyn for discussions and suggestions.
\end{acknowledgements}


\bibliographystyle{aa}

\bibliography{perseus-mosaic}



\appendix

\section{Figures}

\sectionmark{Figures}

\begin{figure*}
\centering
\includegraphics[width=0.97\textwidth]{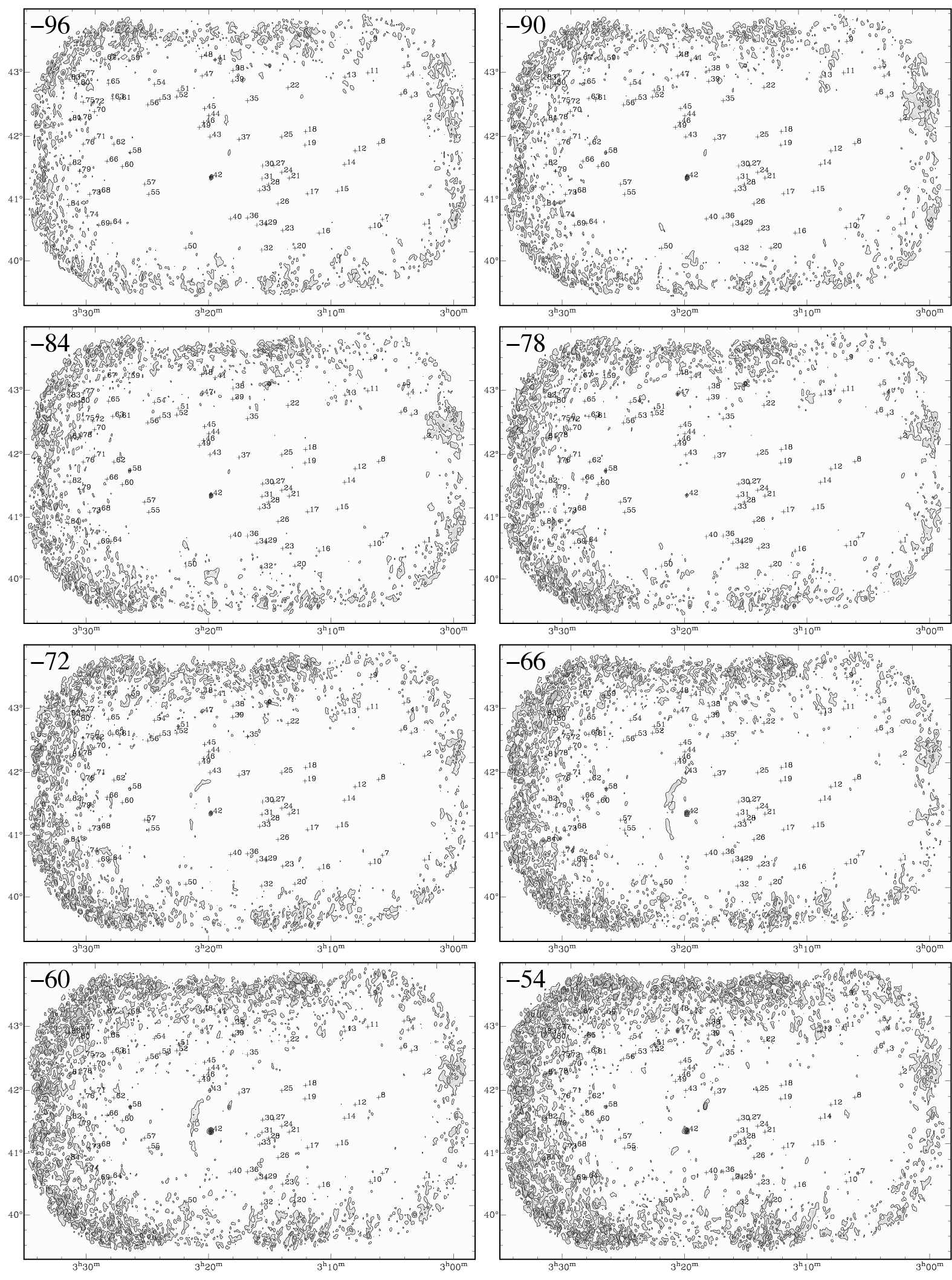}
\caption{Most interesting frames from the RM-cube. The contour
levels are 3, 6, 12, 24,... mJy~beam$^{-1}$~rmtf$^{-1}$ \modified{(1.4, 2.8,
5.5,\ldots~K~rmtf$^{-1}$)}. The Faraday depth in rad~m$^{-2}$ is
indicated in the top left corner of each frame. The HPBW is
$2\arcmin\times3\arcmin$.}
\label{brentjens_perseusmosaic_fig:rmcube-1}
\end{figure*}

\begin{figure*}
\centering
\includegraphics[width=0.97\textwidth]{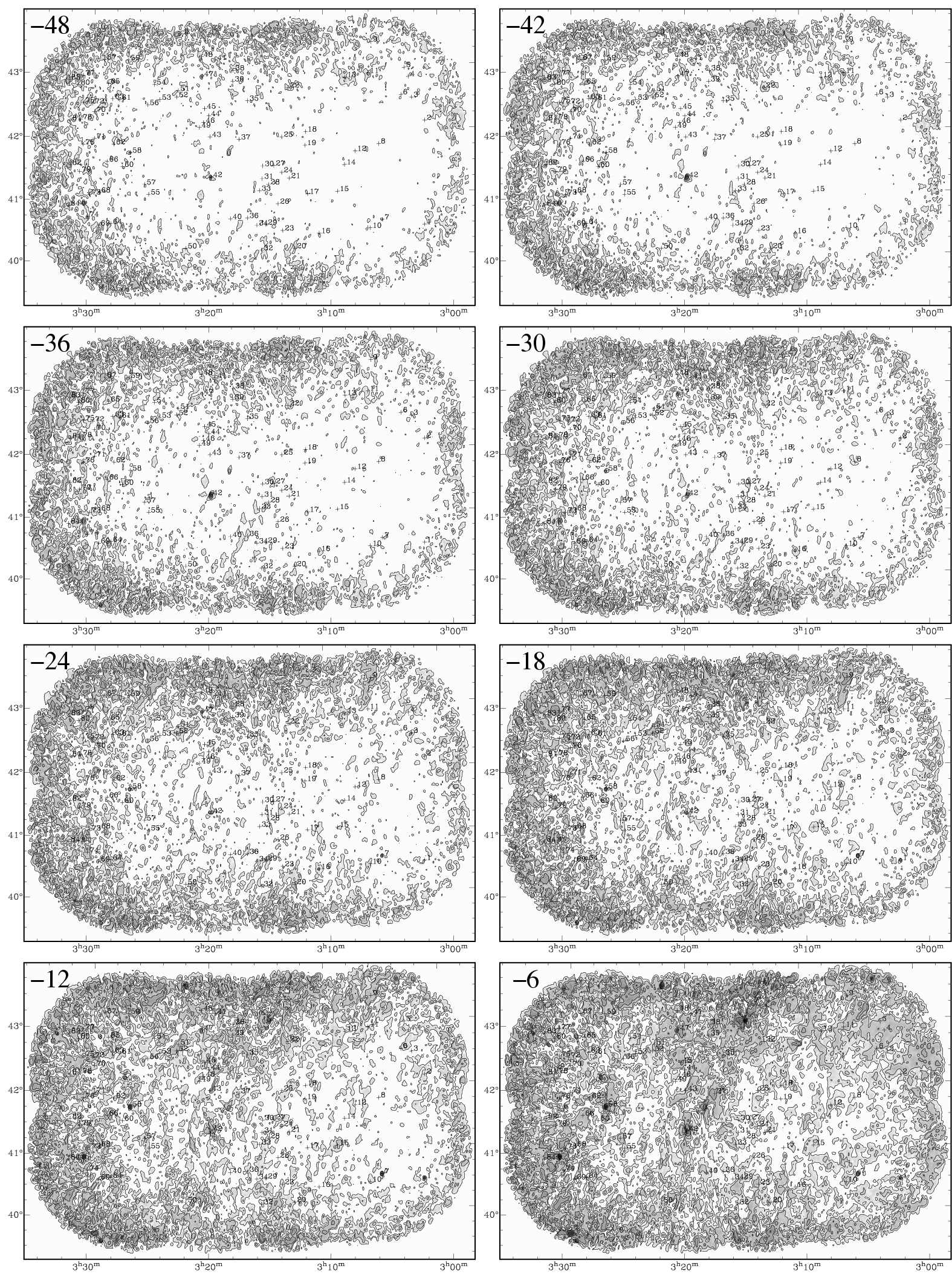}
\caption{Continued from Fig.~\ref{brentjens_perseusmosaic_fig:rmcube-1}.}
\label{brentjens_perseusmosaic_fig:rmcube-2}
\end{figure*}

\begin{figure*}
\centering
\includegraphics[width=0.97\textwidth]{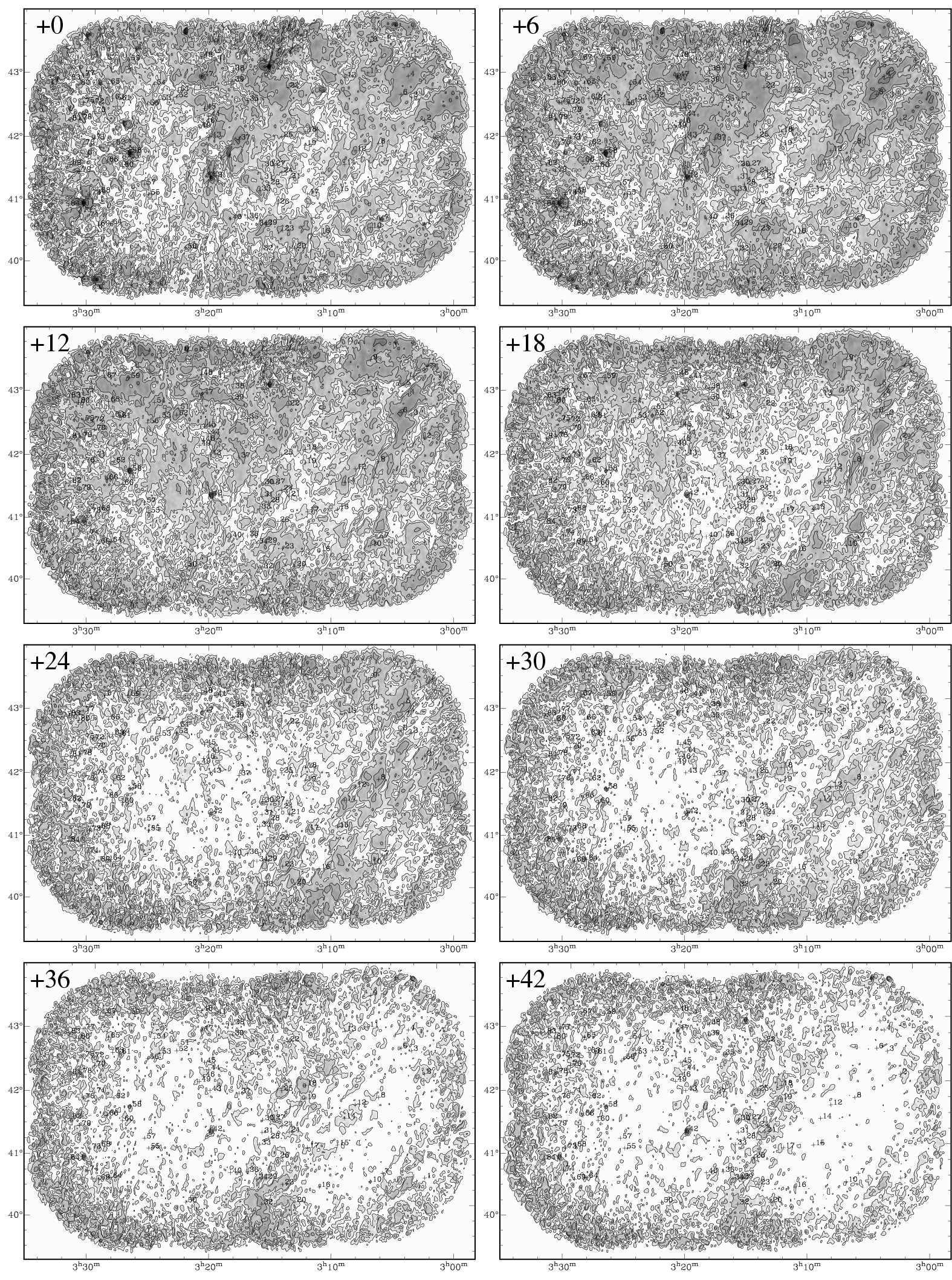}
\caption{Continued from Fig.~\ref{brentjens_perseusmosaic_fig:rmcube-2}.}
\label{brentjens_perseusmosaic_fig:rmcube-3}
\end{figure*}

\begin{figure*}
\centering
\includegraphics[width=0.97\textwidth]{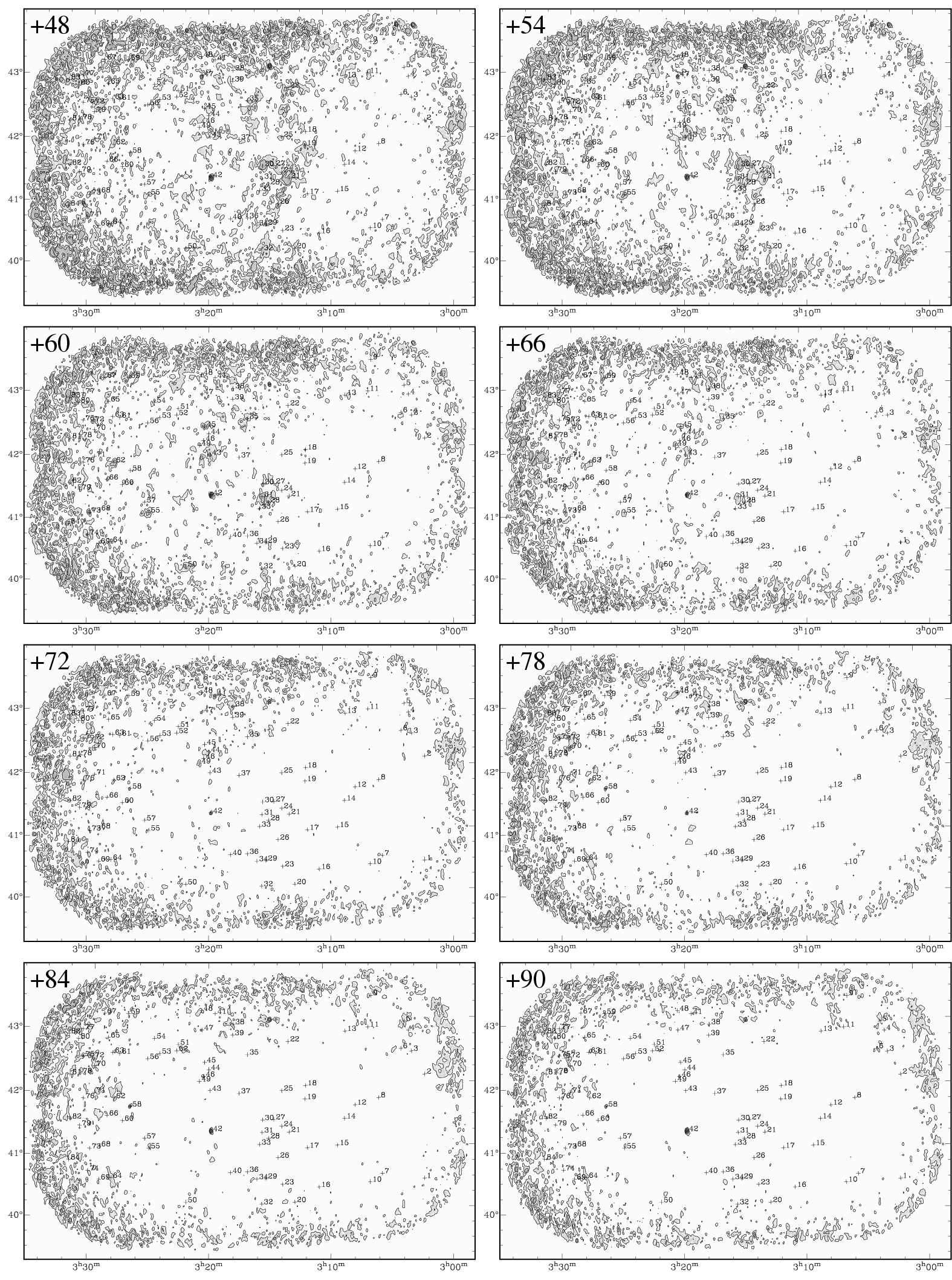}
\caption{Continued from Fig.~\ref{brentjens_perseusmosaic_fig:rmcube-3}.}
\label{brentjens_perseusmosaic_fig:rmcube-4}
\end{figure*}

\begin{figure*}
\centering
\includegraphics[width=0.9\textwidth]{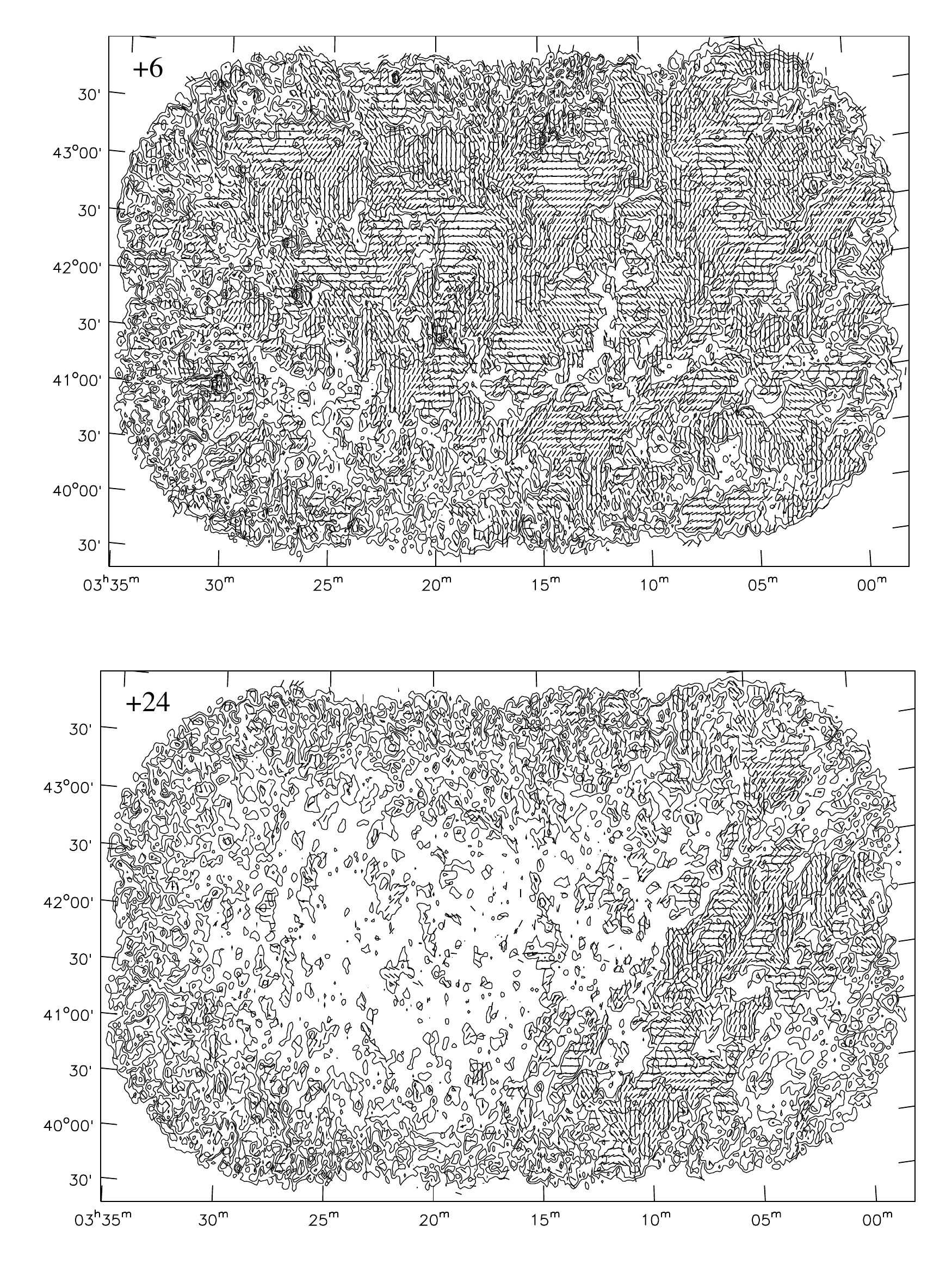}
\caption{Polarization vectors (electric field) at 350.22~MHz overlaid
  on contour maps of the absolute value of the Faraday dispersion
  function. Contours are drawn at $+3$, $+6$ ,$+12$,
  etc. ~mJy~beam$^{-1}$~rmtf$^{-1}$. The polarization vectors are
  drawn if \modified{$|F(\phi)|$ is larger than 5 times the local observed RMS
  at $|\phi| > 100$~rad~m$^{-2}$}. The Faraday depth in rad~m$^{-2}$ is
  plotted in the top left corner of each image.}
\label{brentjens_perseusmosaic_fig:polangles-1}
\end{figure*}

\begin{figure*}
\centering
\includegraphics[width=0.9\textwidth]{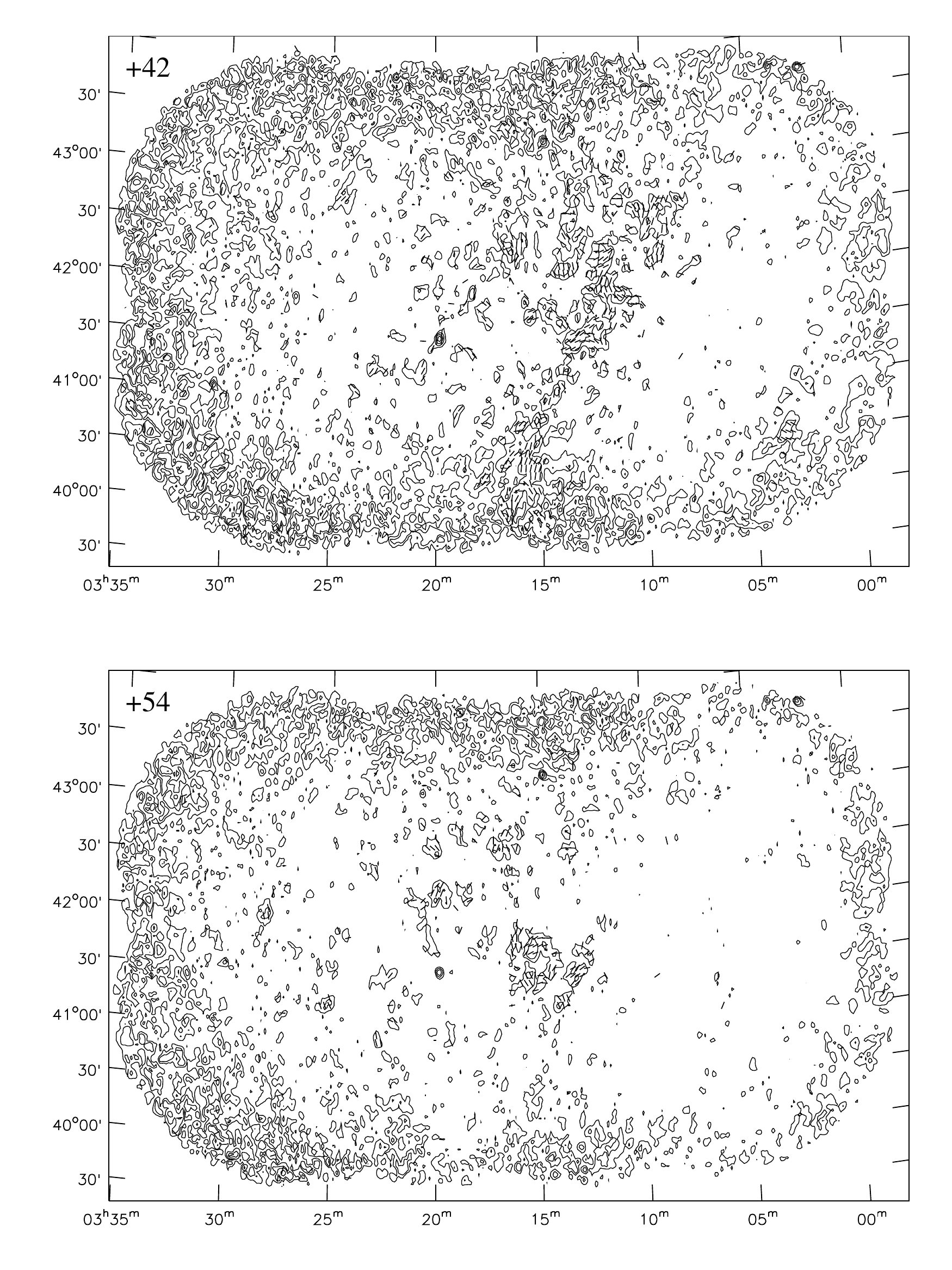}
\caption{Continued from Fig.~\ref{brentjens_perseusmosaic_fig:polangles-1}.}
\label{brentjens_perseusmosaic_fig:polangles-2}
\end{figure*}

\end{document}